\documentclass[11pt]{article}
\usepackage[utf8]{inputenc}
\usepackage{graphicx}
\usepackage{amsfonts}
\usepackage{amsmath,amsthm}
\newtheorem{theorem}{Theorem}

\newtheorem{definition}{Definition}
\newtheorem{lemma}{Lemma}
\newtheorem{algorithm}{Algorithm}
\usepackage{bbm}
\usepackage{amssymb}
\allowdisplaybreaks
\usepackage{color}
\usepackage{float}
\usepackage{url}
\title{Subset-Reach Estimation in Cross-Media Measurement}
\author{{Chenwei Wang, Jiayu Peng, Rieman Li, Ying Liu}\\
{\small Google LLC, US}\\
{\small E-mail:~\{chenweiwang, jiayupeng, riemanli, yingliug\}@google.com}}
\date{June 1, 2023}

\begin{document}

\newcommand{\jiayu}[1]{{\it \color{red} \#jiayu #1}}

\maketitle

\begin{abstract}
We propose two novel approaches to address a critical problem of reach measurement across multiple media -- how to estimate the reach of an unobserved subset of buying groups (BGs) based on the observed reach of other subsets of BGs. Specifically, we propose a model-free approach and a model-based approach. The former provides a coarse estimate for the reach of any subset by leveraging the consistency among the reach of different subsets. Linear programming is used to capture the constraints of the reach consistency. This produces an upper and a lower bound for the reach of any subset.
The latter provides a point estimate for the reach of any subset. The key idea behind the latter is to exploit the conditional independence model. In particular, the groups of the model are created by assuming each BG has either high or low reach probability in a group, and the weights of each group are determined through solving a non-negative least squares (NNLS) problem. In addition, we also provide a framework to give both confidence interval and point estimates by integrating these two approaches with training points selection and parameter fine-tuning through cross-validation. Finally, we evaluate the two approaches through experiments on synthetic data.
\end{abstract}

\newpage
\tableofcontents

\newpage
\parskip 1ex

\section{Introduction}\label{sec:intro}

\begin{figure}[!h]
\centering
\includegraphics[scale=0.5]{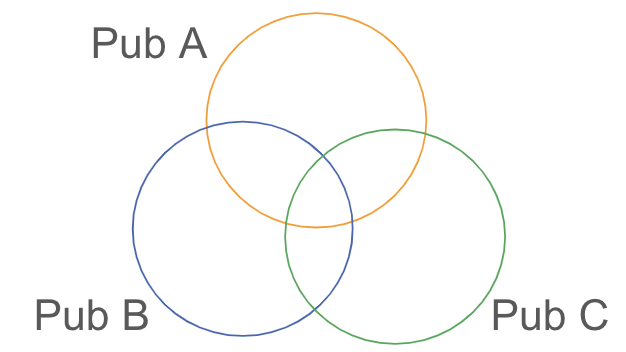}
\vspace{-0.1in}
\caption{Three-publisher reach Venn diagram}
\label{fig:venn_diagram_3set}
\end{figure}

An essential goal of brand advertising is to reach as many potential customers as possible. For instance, as shown in Figure \ref{fig:venn_diagram_3set}, consider an advertiser launched an ad campaign on 3 publishers. Suppose each publisher tells, from their raw data, that it reached 100000 users. Then the advertiser may ask whether the branding budget was effectively spent on, e.g., Pub $C$? To answer this question, we can compute the number of unique users reached by Pub $C$ only, i.e., the incremental reach from the subset of publishers $\{A, B\}$ to the subset $\{A,B,C\}$. Without extra information, the answer can be any number from 0 to 100000, which is a wide range. In the current practice, advertisers often have little clue about where the actual incremental reach falls in this wide range -- and thus little clue about whether their branding budget was effectively spent. More generally, advertisers often want to compute the full reach Venn diagram -- the number of unique users reached across any subset of the publishers, or in other words, the reach of any subset.

A $P$-publisher reach Venn diagram includes $2^P$ subsets. As the number of publishers increases, and especially in a privacy and compute constrained environment, it can be infeasible to compute all possible combinations. Thus, instead of the full reach Venn diagram, we often can only observe a partial reach Venn diagram, i.e., the reach of some subsets. Based on the reach of some subsets, we want to estimate the reach of any other subset. Therefore, in this paper, we propose the methodologies to estimate the reach of any subset.

In practice, the concept of a publisher can also be extended to a buying group (BG). A BG can be a publisher, or a sub-platform of a publisher, or a group of events of interest defined by the advertisers (e.g., a sub-campaign with a specific configuration like special targeting or the campaign within a sub-time-window such as the first week). Thus, the phrase of \emph{subset-reach estimation} used in this paper refers to \emph{estimating the reach of a subset of BGs}. 


In the rest of this section, we introduce an example of incremental reach estimation under any permutation of BGs to show the critical need for estimating the reach of any subset. At the end, we introduce the organization of this paper.

\subsection*{Incremental Reach Under a Permutation of BGs}

Consider an example with $P=5$ BGs denoted by $G_1, G_2, \cdots, G_5$. Further suppose we observe that the reach of each BG is $100000$ and the union reach of all 5 BGs is $336160$. Then given the incremental sequence $G_1\rightarrow G_2\rightarrow G_3\rightarrow G_4\rightarrow G_5$, what is the estimate of the union reach of BGs $G_1,G_2$, that of BGs $G_1,G_2,G_3$, and that of BGs $G_1,G_2,G_3,G_4$? 

\begin{figure}[!t]
\centering
\includegraphics[scale=0.58]{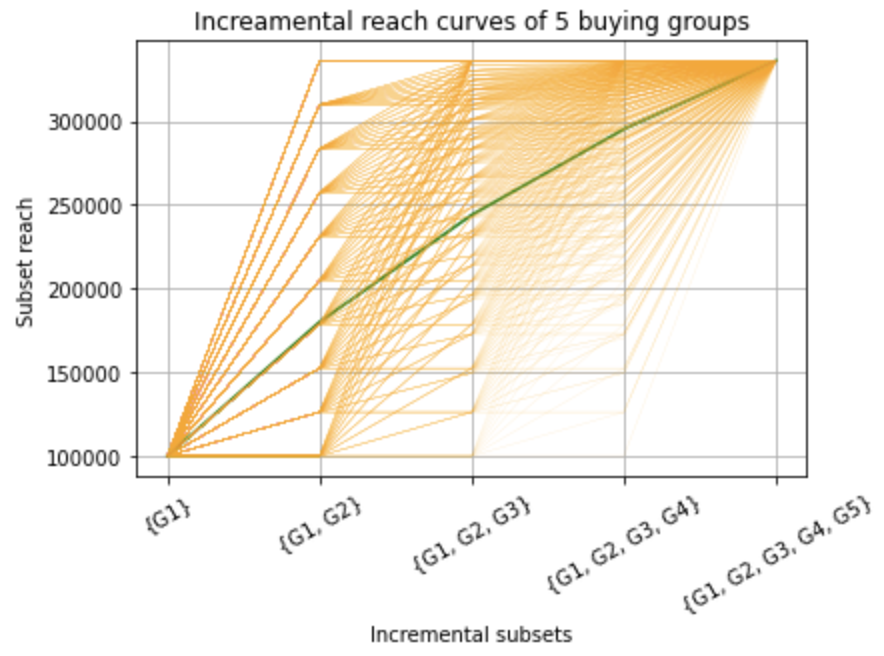}
\vspace{-0.2in}
\caption{An example of incremental reach curves for five BGs.}
\label{fig:incremenral_reach_problem}
\end{figure}

Without extra information, the reach of the subsets, denoted by $R(\{G_1,G_2\})$, $R(\{G_1,G_2,G_3\}\!)$, and $R(\{G_1,G_2,G_3,G_4\})$, respectively, can actually take on any value between $100000$ and $336160$. This is because they are all bounded by the single-BG reach and the all-BGs-union reach, while satisfying $R(\{G_1, G_2\}) \leq R(\{G_1, G_2,G_3\}) \leq R(\{G_1,G_2,G_3,G_4\})$ due to the reach consistency. As shown in Figure \ref{fig:incremenral_reach_problem}, the incremental reach curve can be any orange-colored curve. However, compared to the green-colored ground truth, the estimation range $[100000, 336160]$ is wide. 

To reduce the estimation uncertainty, we need to measure the reach of more subsets. Can we directly measure the reach of the three subsets in the incremental sequence $G_1\rightarrow G_2\rightarrow G_3\rightarrow G_4\rightarrow G_5$? While the answer is yes for this specific sequence, what if the advertiser also wants to know the incremental reach for another sequence, say, $G_2\rightarrow G_3\rightarrow G_5\rightarrow G_1\rightarrow G_4$? Then similar to the initial problem, we observe the reach of only some subsets in this incremental sequence. Again, as the number of queries that an advertiser can afford is limited, we cannot measure the reach of all subsets in the incremental sequence under all possible permutations of the BGs.

Back to $G_1\rightarrow G_2\rightarrow G_3\rightarrow G_4\rightarrow G_5$, suppose we observe the reach of five additional subsets: $R(\{G_1, G_5\})= R(\{G_1, G_4\})=180000$, $R(\{G_2, G_3, G_4\})$ $=R(\{G_1, G_2, G_4\})$ $=244000$, and
$R(\{G_2, G_3, G_4, G_5\}) = 295200$. With this extra information, can we have a narrower range for the estimates of $R(\{G_1,G_2\})$, $R(\{G_1,G_2,G_3\})$, and $R(\{G_1,G_2,G_3,G_4\})$ than $[100000, 336160]$? Naively, this appears difficult because these five additional subsets are not contained in the incremental sequence. Nevertheless, it is straightforward that both $R(\{G_1, G_4\})$ and $R(\{G_2, G_3, G_4\})$ cannot exceed $R(\{G_1,G_2, G_3, G_4\})$ owing to the reach consistency. Hence, we can at least narrow the range of the estimation of $R(\{G_1,G_2, G_3, G_4\})$ down to $[244000, 336160]$. 

The observation above indicates that even if a subset is not contained in the incremental sequence, the reach of that subset may still provide useful information for estimating the reach of a subset contained in that sequence. Besides the incremental reach, in more general use cases, it is critical to understand how to utilize the observed reach of some subsets (of BGs) to estimate the reach of an unobserved subset (of BGs).

The rest of this paper is organized as follows. Sec. \ref{sec:problem_solution} introduces the problem formulation and a brief overview of our proposed new approaches. The proposed approaches include both a model-free and a model-based approach, which are described in detail and exemplified in Sec. \ref{sec:model_free} and Sec. \ref{sec:DMM}, respectively. In Sec. \ref{sec:framework}, we integrate the two approaches into a framework, which incorporates adaptively selecting more training points, fine-tuning the model parameter, and suggesting a method for calculating the error bar. To evaluate our proposed approaches, we provide two experiments in Sec. \ref{sec:simulation} by using synthetic data. Finally, we conclude this paper in Sec. \ref{sec:conclusion}.

\section{Problem Description and Our Proposed Approaches}\label{sec:problem_solution}

In this section, we begin with the problem formulation and provide a brief overview of our proposed new approaches. We also summarize the glossary and the notations used throughout this paper.

\subsection{Problem Description}

Consider $P$ BGs, denoted by $G_1,~G_2,~\cdots,~G_P$. Suppose we observe the reach of $n$ subsets of the $P$ BGs, which include at least $P$ single-BG reach and the union reach of all the $P$ BGs, and/or the reach of some other subsets. We are interested in addressing the following three questions: 
\begin{enumerate}
    \item How to examine if the reach of these $n$ observed subsets satisfy the reach consistency?
    
    \item How to estimate the reach of an unobserved subset? 
    
    \item How to evaluate the estimated reach of an unobserved subset?
\end{enumerate}

Note that among the three related questions above, the most critical question is the second.

\subsection{Overview of Our Proposed Approaches}

In this paper, we propose a (statistical) model-free approach and a (statistical) model-based approach to answer all the three questions posed above. In particular, we provide an overview of our solutions in this section and introduce them in detail in the rest of this paper. 

In the model-free approach, we leverage the consistency among the reach Venn diagram. Using the reach consistency as the constraints, we develop a linear programming solver. Using this solver, we address the first question by transforming the reach-consistency check problem into a linear programming problem that can be efficiently solved. The solution is specified in Theorem \ref{theorem:detecting_consistency} in Sec. \ref{sec:model_free}. 

Additionally, we can bound the reach of any unobserved subset with minor adjustments to the linear programming solver. Thus, we resolve the second question by providing a $100\%$-confidence interval estimate. This solution is described in Theorem \ref{theorem:bound_subset_reach} in Sec. \ref{sec:model_free}. The main idea behind the model-free approach is that the reach of any subset can be broken down into the reach of the primitive regions belonging to that subset. The possible reach range of any subset, therefore, can be determined through the imposed constraints of the reach of the primitive regions in the linear programming solver.

In the model-based approach, we propose an algorithm to fit the reach of the observed subsets to a conditional independence model. While the concept of the conditional independence model has been well investigated in some use cases \cite{Shen2019ConditionalIA}, a challenge arises when the problem is non-parametric or semi-parametric, e.g., the number of groups can be unlimited. To deal with this challenge, we divide the universe, i.e., the set of all the users, into $2^P$ mutually-exclusive activity segments by assuming that each BG has low- or high-reach probability in a segment. Then in each segment, we apply an independence-model assumption that the probability of a user being reached by any BG is independent of the others. Finally, the reach of an observed subset is approximated by a linear combination of its reach in each segment, and the weights are determined through solving an NNLS problem. With this approach, we answer the second question by providing a point estimate. This solution is present in Algorithm \ref{alg:model_fit} in Sec. \ref{sec:DMM}. 

\begin{figure}[!t]
\centering
\includegraphics[scale=0.5]{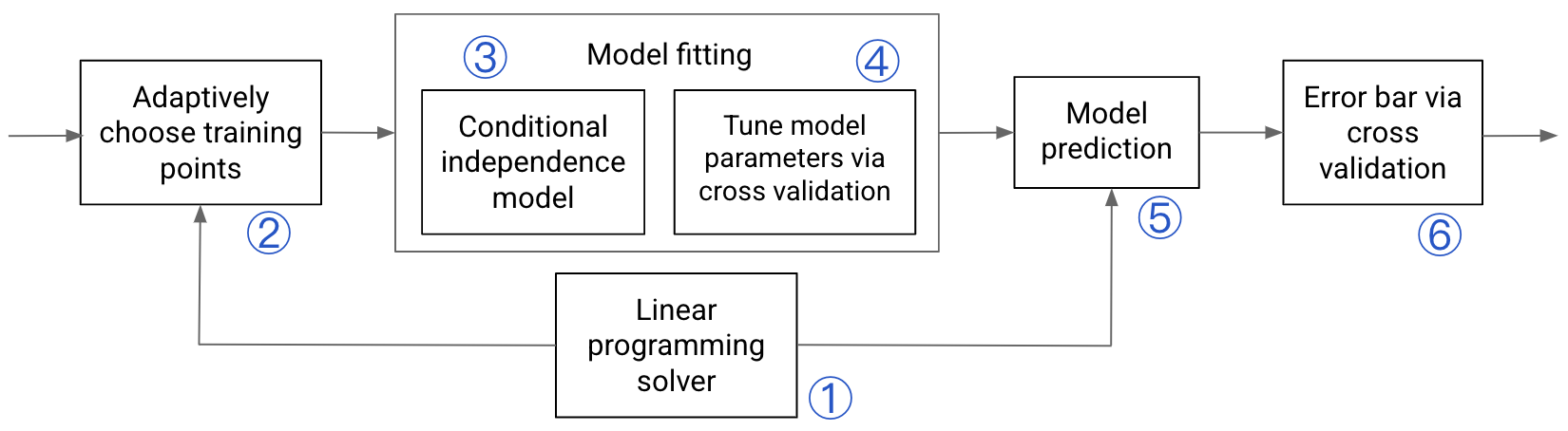}\vspace{-0.1in}
\caption{The framework of the proposed reach modeling to estimate the reach an unobserved subset}
\label{fig:framework}
\end{figure}

Finally, we propose a framework depicted in Figure \ref{fig:framework} that integrates the two approaches described above to give both a confidence interval and a point estimate of the reach of any unobserved subset. In addition, this framework includes adaptively determining which subset to measure the reach for training the proposed model in Algorithm \ref{alg:model_fit}, fine-tuning the model parameter via cross-validation, and suggesting a method to define an error bar for the evaluation purpose. These topics are thoroughly discussed in Sec. \ref{sec:framework}.

\subsection{Glossary and Notations}

We summarise the phrases frequently used throughput this paper in Table \ref{tab:glossary}. 
\begin{table}[H]
\centering
\caption{Glossary}
\vspace{0.5em}
\begin{tabular}{p{8em}p{32em}}
    \hline\hline
    Glossary & Description \\ \hline
    BG & A BG (buying group) can be a publisher, or a sub-platform of a publisher, or a group of events of interest defined by the advertisers. \\ 
    Conditional independence & The universe consists of multiple mutually-exclusive groups, and within each group, the probability of a user being reached by one BG is independent of the others.\\ 
    Venn diagram & A Venn diagram uses overlapping circles or other shapes to illustrate the logical relationships between two or more sets of items. \\
    Primitive region & an indivisible region of a Venn diagram\\
    Reach of a subset & the number of users reached by a subset of BGs\\
    Training points & the reach of the observed subsets\\
    NNLS & non-negative least squares\\
    \hline\hline
\end{tabular}\label{tab:glossary}
\end{table}

For brevity, we also summarize the notations used throughout this paper in Table \ref{tab:notations}.
\begin{table}[H]
\centering
\caption{Notations}
\vspace{0.5em}
\begin{tabular}{p{5em}p{35em}}
    \hline\hline
    Notations  & Description \\ \hline
    $P$ & the number of BGs\\ 
    $\mathcal{U}$ & the universe, i.e., the set of all the users\\ 
    $U$ & the universe size, i.e., the number of users in $\mathcal{U}$ \\ 
    $G_i$ & the $i^{th}$ BG where $i=1,\cdots,P$ \\
    $\mathcal{G}_i$ & the subset of the users reached by BG $G_i$ \\ 
    $\mathcal{S}$ & a subset of BGs. $\mathcal{S}$ can be a single-BG set, e.g., $\{G_1\}$ or a union set of multiple BGs, e.g., $\{G_1,G_2\}$. For $P$ BGs, there are a total of $2^P-1$ non-empty subsets.\\ 
    $R(\mathcal{S})$ & the reach of the subset $\mathcal{S}$ of BGs, i.e., the number of users reached by at least one BG in $\mathcal{S}$. \\
    $r(\mathcal{S})$ & the proportion of the users in the universe that are reached by at least one BG in $\mathcal{S}$ and is expressed as $r(\mathcal{S})=R(\mathcal{S})/U$. It also represents the probability of a user being reached by at least one BG in $\mathcal{S}$.\\ 
    $a$, ${\bf a}$, ${\bf A}$ & a scalar, a column vector, and a matrix, respectively\\
    $|\cdot|$ & the cardinality of a set or the absolute value of a scalar \\
    $\|\cdot \|_p$ & the $L_p$ norm of the vector\\
    $(\cdot)^T$ & the transpose operator of a vector or a matrix \\
    $\cup$, $\cap$, $\setminus$ & the set union, the set intersection, and set difference operators, respectively \\
    ${\bf a} \succeq {\bf b}$ & for two vectors ${\bf a}$ and ${\bf b}$ with equal size, each entry of the ${\bf a}$ is greater than or equal to the corresponding entry of ${\bf b}$. \\
    $\textrm{bin2dec}(\cdot)$ & a function that translates a binary representation to its decimal representation, e.g.,  $\textrm{bin2dec}(110)=6$ \\ \hline\hline
\end{tabular}\label{tab:notations}
\end{table}

Note that given a subset of BGs, e.g., $\mathcal{S}=\{G_1,G_2\}$, the reach of that subset, based on the notations defined in Table \ref{tab:notations}, is expressed as $R(\{G_1,G_2\})\triangleq |\mathcal{G}_1\cup \mathcal{G}_2|$, the cardinality of the (union) subset $\mathcal{G}_1\cup \mathcal{G}_2$. Thus, for brevity and notation consistency, when no ambiguity is caused, we use $\mathcal{S}$ to denote a subset of BGs (e.g., $\{G_1,G_2\}$) or a subset of users reached by at least one of those BGs (e.g., $\mathcal{G}_1\cup \mathcal{G}_2$), interchangeably. In addition, for any set operation expression involving union, intersection, and set difference operators, e.g., $\mathcal{G}_1\setminus \mathcal{G}_2$, we also use $R(\mathcal{G}_1\setminus \mathcal{G}_2)$ to denote the reach of that expression, i.e., the number of the users reached by BG $G_1$ but not by BG $G_2$. 


\section{A Model-free Approach: Bound the Reach of a Subset Based on Venn Diagram Consistency}\label{sec:model_free}

In this section, we propose a model-free approach to bound the reach of any unobserved subset. We begin with an example to demonstrate how to examine the reach of subsets satisfy the reach consistency. Then we develop our solution by designing a linear programming solver based on reach Venn diagram. Finally, by using this solver with minor adjustments, we can estimate the reach of any subset by providing an upper and a lower bounds. 

\subsection{Consistency Among a Reach Venn Diagram} 

Consider three BGs $G_1$, $G_2$, and $G_3$. Suppose we observe each single-BG reach $R(\mathcal{G}_i)=3000$ for $i=1,2,3$, the all-BGs-union reach $R(\mathcal{G}_1\cup \mathcal{G}_2\cup \mathcal{G}_3) = 7000$, and $R(\mathcal{G}_2 \cup \mathcal{G}_3) = 5000$. Then we want to estimate the unobserved $R(\mathcal{G}_1 \cup \mathcal{G}_3)$. Consider the estimations of two models in the following:
\begin{itemize}
    \item Model 1 estimates $\hat{R}(\mathcal{G}_1 \cup \mathcal{G}_3)=3500$. Is this estimation feasible? The answer is no, because $R(\mathcal{G}_1\cup \mathcal{G}_2\cup \mathcal{G}_3) = 7000$ is larger than $ R(\mathcal{G}_1 \cup \mathcal{G}_3) + R(\mathcal{G}_2)=6500$, which contradicts the triangle inequality $R(\mathcal{G}_1\cup \mathcal{G}_2\cup \mathcal{G}_3) \leq R(\mathcal{G}_1 \cup \mathcal{G}_3) + R(\mathcal{G}_2)$.
    
    \item Model 2 estimates $\hat{R}(\mathcal{G}_1 \cup \mathcal{G}_3)=4000$. Again, is it feasible? Now since the triangle inequality above is satisfied, this estimation seems feasible. However, it is still not, because $R((\mathcal{G}_1\setminus \mathcal{G}_3) \cup (\mathcal{G}_2\setminus \mathcal{G}_3)) =  R(\mathcal{G}_1\cup \mathcal{G}_2\cup \mathcal{G}_3) - R(\mathcal{G}_3) = 4000$ is larger than $ R(\mathcal{G}_1\setminus \mathcal{G}_3) + R(\mathcal{G}_2\setminus \mathcal{G}_3)=R(\mathcal{G}_1\cup \mathcal{G}_3) -R(\mathcal{G}_3)+ R(\mathcal{G}_2\cup \mathcal{G}_3)-R(\mathcal{G}_3)= 3000$, which contradicts another triangle inequality $R((\mathcal{G}_1\setminus \mathcal{G}_3) \cup (\mathcal{G}_2\setminus \mathcal{G}_3))\leq  R(\mathcal{G}_1\setminus \mathcal{G}_3) + R(\mathcal{G}_2\setminus \mathcal{G}_3)$.
\end{itemize}

The example above implies that the possible values of $R(\mathcal{G}_1 \cup \mathcal{G}_3)$ constitute a feasible region that is bounded by the underlying inequalities of the reach of other subsets. If $R(\mathcal{G}_1 \cup \mathcal{G}_3)$ takes a value outside of that region, $R(\mathcal{G}_1 \cup \mathcal{G}_3)$ is inconsistent with the reach of other subsets. With this intuition, we define the reach consistency in the following.

\begin{definition}\label{def:consistency}
(Reach consistency)
The reach of any two subsets $\mathcal{S}_1$, $\mathcal{S}_2$ are consistent if and only if they satisfy (Monotonicity) $R(\mathcal{S}_1 \cup \mathcal{S}_2) \geq  R(\mathcal{S}_i),~i=1,2$ and (Subadditivity) $R(\mathcal{S}_1 \cup \mathcal{S}_2) \leq R(\mathcal{S}_1) + R(\mathcal{S}_2)$.
\end{definition}

From the example above, it can be seen that examining the monotonicity and subadditivity of the reach of any subsets is not straightforward for the case with even $P=3$ BGs, let alone for $P\geq 3$. Thus, it is necessary to develop a systematic approach that is computation-efficient to validate the reach consistency. 

Before proceeding into our proposed approach, we first introduce the following preliminaries.

\begin{definition} \label{def:pr}
(Primitive regions)
Refer to each indivisible region of a Venn diagram as a primitive region. A $P$-BGs reach Venn diagram has $2^P$ primitive regions. Denote them by $\mathcal{R}_{x_1\cdots x_P}=\cap_{\forall x_i=1}~\mathcal{G}_i\setminus \cup_{\forall x_i=0}~ \mathcal{G}_i$ where $x_i\in\{0,1\},~i=1\cdots,P$.
\end{definition}

{\it Example}: Consider $P=2$ BGs $G_1$ and $G_2$. A $2$-BGs reach Venn diagram has $2^P=2^2=4$ primitive regions: $\mathcal{G}_1\setminus \mathcal{G}_2$, $\mathcal{G}_2\setminus \mathcal{G}_1$, $\mathcal{G}_1\cap \mathcal{G}_2$ indicating the set of users reached by $G_1$ only, $G_2$ only, both $G_1$ and $G_2$, respectively, and $\mathcal{U}\setminus (\mathcal{G}_1\cup \mathcal{G}_2)$ representing the users not reached by any BG. For brevity, denote them by $\mathcal{R}_{10}$, $\mathcal{R}_{01}$, $\mathcal{R}_{11}$, and $\mathcal{R}_{00}$, respectively. 

\begin{lemma}\label{lemma:sum_pr}
The reach of any subset is the sum of the reach of all the primitive regions belonging to that subset. 
\end{lemma}

{\it Proof}: Denote by $\mathcal{S}$ a subset and by $\mathcal{R}_{x_1\cdots x_P}$'s the $2^P$ primitive regions of the $P$-BGs reach Venn diagram. Then we have
\begin{eqnarray}
R(\mathcal{S})=R(\cup_{\forall \mathcal{R}_{x_1\cdots x_P}\subseteq \mathcal{S}}~\mathcal{R}_{x_1\cdots x_P})=\sum_{\forall \mathcal{R}_{x_1\cdots x_P}\subseteq \mathcal{S}} R(\mathcal{R}_{x_1\cdots x_P})
\end{eqnarray}
where the second equality holds because all the primitive regions are mutually exclusive and $R(\cdot)$ is a linear counting function. \qed

\begin{lemma}\label{lemma:subset_to_pr}
In a $P$-BGs reach Venn diagram, for any subset $S$, there exists a $2^P\times 1$ column vector ${\bf b}_{\mathcal{S}}$ so that
$R(\mathcal{S})={\bf b}_{\mathcal{S}}^T{\bf x}$, where ${\bf x}=[R(\mathcal{R}_1),\cdots,R(\mathcal{R}_{2^P})]^T$ is another $2^P\times 1$ column vector stacking the reach of all the $2^P$ primitive regions defined in Definition \ref{def:pr}. The $j^{th}$ entry of ${\bf b}_{\mathcal{S}}$ for $j=1,\cdots,2^P$ is given by $\mathbbm{1}_{\mathcal{R}_j}(\mathcal{S})$, an indicator function equal to one if $\mathcal{R}_j\subset \mathcal{S}$ and zero otherwise. 
\end{lemma}

{\it Proof}: The existence and construction of the vector ${\bf b}_{\mathcal{S}}$ for any subset $\mathcal{S}$ directly follows Lemma \ref{lemma:sum_pr} in the vector form. 
\qed

\begin{lemma}\label{lemma:consistency_to_lp}
A combination of the reach of multiple subsets is consistent if and only if the reach of each subset can be expressed as a partial-sum of the same non-negative reach of the $2^P$ primitive regions.
\end{lemma}

{\it Proof:} This lemma directly follows Definition \ref{lemma:subset_to_pr} in the compact matrix form. \qed

\subsection{Validate the Reach Consistency Through Linear Programming}

As Lemma \ref{lemma:consistency_to_lp} implies, detecting the consistency among the reach of multiple subsets is equivalent to verifying if the reach of each primitive region is non-negative. This can be accomplished by linear programming. Based on this idea, our approach is stated through the following theorem.

\begin{theorem}\label{theorem:detecting_consistency}
Consider $P$ BGs and the reach of $n$ subsets $R(\mathcal{S}_i)$'s for $i=1,\cdots,n$. They are \emph{consistent} if and only if the following linear programming has a non-negative solution:
\begin{eqnarray}
\arg&&\max_{t} ~~t\\
\textrm{s.t.}&&\left\{
\begin{array}{ll}
   R(\mathcal{S}_i)={\bf b}_{\mathcal{S}_i}^T{\bf x}  &  ~~~~i=1,\cdots,n,\\
    t\leq x_j & ~~~~j=1,\cdots,2^P,
\end{array}\right.\label{eqn:consistency_lp_constraint}
\end{eqnarray}
where ${\bf b}_{\mathcal{S}_i}$'s are defined in Lemma \ref{lemma:subset_to_pr}, and ${\bf x}=[x_1,x_2,\cdots,x_{2^P}]^T$ represents the reach of the $2^P$ primitive regions (the order does not affect).
\end{theorem}

{\it Proof:} The proof is relatively straightforward. Specifically, if the solution $t^*\geq 0$, each entry of the vector ${\bf x}$ is non-negative, thus indicating the reach consistency. On the other hand, if the reach consistency is satisfied, each entry of ${\bf x}$ must be non-negative. Thus, $t^*$ is given by the minimum of all the entries of ${\bf x}$, which implies $t^*=\min(x_1,\cdots,x_{2^P})\geq 0$. \qed 

To apply Theorem \ref{theorem:detecting_consistency}, we first need to specify the linear combination vector ${\bf b}_{\mathcal{S}_i}$ for each subset $\mathcal{S}_i$ and then solve the formulated linear-programming problem. If the solution $t^*\geq 0$, $R(\mathcal{S}_i)$'s are consistent; otherwise, they are inconsistent.

\subsection{Bound the Reach of a Subset Through Linear Programming}

Besides detecting the reach consistency, we can also employ linear programming to bound the reach of any unobserved subset. This is because the constraints in (\ref{eqn:consistency_lp_constraint}) have already enforced a collection of limitations on the reach of primitive regions, which in turn can be used to characterize the reach of an unobserved subset. As a result, linear programming is a natural tool to discover the reach boundaries while maintaining the reach consistency. The main result is introduced through the following theorem.   


\begin{theorem}\label{theorem:bound_subset_reach}
Suppose we observe the reach of $n$ subsets $R(\mathcal{S}_i)$ for $i=1,\cdots,n$, and they are consistent. Given an unobserved subset $\mathcal{S}^*$, without extra information, the tightest upper and lower bounds of $R(\mathcal{S}^*)$, denoted by $\overline{R}(\mathcal{S}^*)$ and $\underline{R}(\mathcal{S}^*)$, respectively, are given by:
\begin{eqnarray}
\overline{R}(\mathcal{S}^*)~~=~~\arg&&\min_{t} t\\
\textrm{s.t.}&&\left\{
\begin{array}{ll}
   R(\mathcal{S}_i)={\bf b}_{\mathcal{S}_i}^T{\bf x}  &  \textrm{for}~~~i=1,\cdots,n,\\
    t={\bf b}_{\mathcal{S}^*}^T{\bf x}, &\\
    {\bf x}\succeq {\bf 0},&
\end{array}\right.
\end{eqnarray}
and
\begin{eqnarray}
\underline{R}(\mathcal{S}^*)~~=~~\arg&&\max_{t} t\\
\textrm{s.t.}&&\left\{
\begin{array}{ll}
   R(\mathcal{S}_i)={\bf b}_{\mathcal{S}_i}^T{\bf x}  &  \textrm{for}~~~i=1,\cdots,n,\\
    t={\bf b}_{\mathcal{S}^*}^T{\bf x}, &\\
    {\bf x}\succeq {\bf 0}.
\end{array}\right.
\end{eqnarray}
\end{theorem}

{\it Proof:} The proof follows the similar idea as that for Theorem \ref{theorem:detecting_consistency}. Specifically, for either the upper bound $t^*=\overline{R}(\mathcal{S}^*)$ or the lower bound $t^*=\underline{R}(\mathcal{S}^*)$, it must exist an ${\bf x}^*\succeq {\bf 0}$, which implies the consistency among the reach Venn diagram. Thus, after projecting ${\bf x}^*$ to ${\bf b}_{\mathcal{S}_i}$, the largest/smallest possible projection automatically turn into the tightest upper/lower bound of $R(\mathcal{S}^*)$.   \qed

With the approach shown in Theorem \ref{theorem:bound_subset_reach}, let us revisit the example in Figure \ref{fig:incremenral_reach_problem} to see how to obtain more accurate incremental reach curves. Given the reach of each single-BG and the all-BGs-union reach, we determine $n=6$ in Theorem \ref{theorem:bound_subset_reach}. For each $\mathcal{S}_i$ equal to 
$\mathcal{G}_i$ for $i=1\cdots,5$ and 
$\cup_{i=1}^5 \mathcal{G}_i$, respectively, we specify the corresponding ${\bf b}_{\mathcal{S}_i}$, the underlying linear combination coefficients of the reach of the primitive regions, based on Lemma \ref{lemma:subset_to_pr}. Then for each of the three unobserved subsets, $\mathcal{G}_1\cup \mathcal{G}_2$, $\cup_{i=1}^3 \mathcal{G}_i$, and $\cup_{i=1}^4 \mathcal{G}_i$, we perform the following three steps sequentially:
\begin{enumerate}
    \item Let $\mathcal{S}^*=\mathcal{G}_1\cup \mathcal{G}_2$ and specify the corresponding ${\bf b}_{S^*}$ based on Lemma \ref{lemma:subset_to_pr}. Apply Theorem \ref{theorem:bound_subset_reach} to obtain the upper bound $\overline{R}(\mathcal{G}_1\cup \mathcal{G}_2)$ and the lower bound $\underline{R}(\mathcal{G}_1\cup \mathcal{G}_2)$. Choose any $\hat{R}(\mathcal{G}_1\cup \mathcal{G}_2) \in [\overline{R}(\mathcal{G}_1\cup \mathcal{G}_2), ~\underline{R}(\mathcal{G}_1\cup \mathcal{G}_2)]$. 
    
    \item Treat the chosen $\hat{R}(\mathcal{G}_1\cup \mathcal{G}_2)$ as the ground truth and increase $n$ by one, i.e., $n=6+1=7$. Let $\mathcal{S}^*=\cup_{i=1}^3 \mathcal{G}_i$ and specify the corresponding ${\bf b}_{S^*}$ based on Lemma \ref{lemma:subset_to_pr}. Then apply Theorem \ref{theorem:bound_subset_reach} to obtain the upper bound $\overline{R}(\cup_{i=1}^3 \mathcal{G}_i)$ and the lower bound $\underline{R}(\cup_{i=1}^3 \mathcal{G}_i)$. Pick any $\hat{R}(\cup_{i=1}^3 \mathcal{G}_i)$ between these two bounds.
    
    \item Also, treat $\hat{R}(\cup_{i=1}^3 \mathcal{G}_i)$ as the ground truth and update $n=7+1=8$. Let $S^*=\cup_{i=1}^4 \mathcal{G}_i$ and specify the corresponding ${\bf b}_{S^*}$ based on Lemma \ref{lemma:subset_to_pr}. Then apply Theorem \ref{theorem:bound_subset_reach} to obtain the upper bound $\overline{R}(\cup_{i=1}^4 \mathcal{G}_i)$ and the lower bound $\underline{R}(\cup_{i=1}^4 \mathcal{G}_i)$. Pick any $\hat{R}(\cup_{i=1}^4 \mathcal{G}_i)$ between the two bounds. Finally, the estimated reach of the three unobserved subsets contained in the incremental sequence is given by $\hat{R}(\cup_{i=1}^2 \mathcal{G}_i)$, $\hat{R}(\cup_{i=1}^3 \mathcal{G}_i)$ and $\hat{R}(\cup_{i=1}^4 \mathcal{G}_i)$. 
\end{enumerate}

\begin{figure}[!t]
\centering
\includegraphics[scale=0.58]{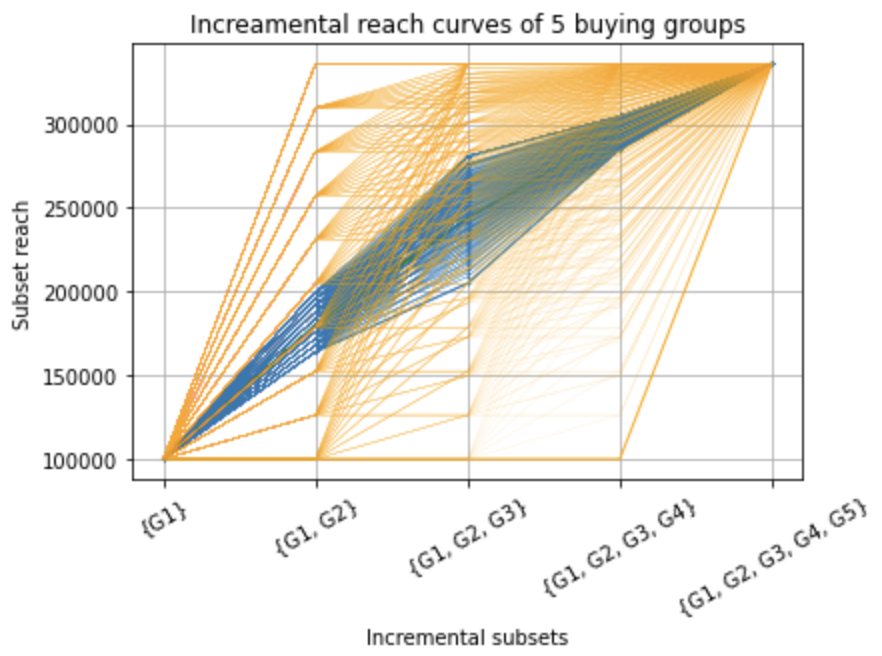}
\caption{An example of incremental reach curves for five BGs.}
\label{fig:incremenral_reach_linear_programming}
\end{figure}

To see the feasible region of the incremental reach curves, we choose a different $\hat{R}(\cdot)$ within its range $[\overline{R}(\cdot), \underline{R}(\cdot)]$ in each step and use it as the ground truth in the next step. As a result, we obtain a family of incremental reach curves, shown as the blue-colored dense curves/region in Figure \ref{fig:incremenral_reach_linear_programming}. It can be seen that the blue-colored curves span a much narrower range than the orange-colored curves, and the boundaries of the blue-colored region are much closer to the ground truth. Meanwhile, the blue-colored region covers the ground truth with $100\%$ probability. Thus, Theorem \ref{theorem:bound_subset_reach} provides a fundamental limit for estimating the reach of a the subset no matter what models/algorithms can be used. 

Note that if we desire to identify only the boundaries of this feasible region, we can directly choose $\hat{R}(\cdot)$ to be $\overline{R}(\cdot)$ and $\underline{R}(\cdot)$, respectively in each step, and find the upper and lower bound of the reach estimation in the next step. Then sequentially connecting all the upper bounds and all the lower bounds, respectively, give the upper and lower boundaries of the feasible region, respectively.

\section{A Model-based Approach: Conditional Independence Model}\label{sec:DMM}

Compared to the model-free approach, modeling is usually more appealing owing to its ability to better characterize the dependencies among multiple BGs and to explore more advanced optimization techniques. In this section, we propose a model-based approach to estimate the reach of an unobserved subset. In particular, by using the reach of observed subsets as the training points, we design a conditional independence model and propose an algorithm to fit the training points to the model. Also, by tuning a parameter of the model, we show that the model can be perfectly fitted to achieve zero training error.
Next, we start from the model description, and then provide examples and insights to interpret the proposed approach.

\subsection{The Proposed Model-based Approach}\label{sec:improved_DMM}

We begin with the following definitions as preliminaries.

\begin{definition}\label{def:binary_strings}
($P$-length binary string)
For $P$ BGs, define a set $\mathcal{I}=\{x_1\cdots x_P~|~x_i\in\{0,1\},~i=1,\cdots,P\}$ where each element of $\mathcal{I}$ is a $P$-bit binary string.
\end{definition}

{\it Remark:} Since each $x_i$ is a binary, there are a total of $2^P$ elements in $\mathcal{I}$.

\begin{definition}\label{def:segment}
(Universe segmentation)
For $P$ BGs, an universe $\mathcal{U}$ consists of $2^P+1$ mutually-exclusive activity segments. In the first $2^P$ segments $\mathcal{U}_{x_1\cdots x_P}$ for each $x_1\cdots x_P\in \mathcal{I}$ defined in Definition \ref{def:binary_strings}, BG $G_i$ has a high probability to reach a user if $x_i=1$ or a low probability to reach a user if $x_i=0$. The last activity segment corresponds to the subset of users in universe that are not reached by any BG.
\end{definition}

{\it Remark:} We will use Definition \ref{def:segment} as an assumption for our model-based approach to assign each user in the universe to one activity segment. The high/low-reach probability will be modeled later in Definition \ref{def:column_z}.

{\it Example:} For $P=3$ and $x_1x_2x_3=110$,
$\mathcal{U}_{110}$ implies an activity segment in which each user is reached by both $G_1$ and $G_2$ with a high probability and by $G_3$ with a low probability.

\begin{definition}\label{def:minterm_maxterm}
(Union subset)
Consider $P$ BGs and $\mathcal{G}_i$, the set of users reached by BG $G_i$, where $i=1,\cdots,P$. For each $x'_1x'_2\cdots x'_P\in \mathcal{I}$ defined in Definition \ref{def:binary_strings}, we define the following union subset
\begin{eqnarray}
\mathcal{S}_{x'_1x'_2\cdots x'_P} \triangleq \cup_{\forall x'_i=1} \mathcal{G}_i.
\end{eqnarray}
If a user belongs to $\mathcal{S}_{x'_1x'_2\cdots x'_P}$, we say that user is reached by the subset $\mathcal{S}_{x'_1x'_2\cdots x'_P}$ for brevity. 
\end{definition}

{\it Remark:} In a $P$-BGs reach Venn diagram, $\mathcal{S}_{x'_1x'_2\cdots x'_P}$ implies the subset of users reached by at least one BG $G_i$ with $x'_i=1$. As a result,  $R(\mathcal{S}_{x'_1x'_2\cdots x'_P})$ represents the (union) reach of the subset of all the BGs $\{G_i~|~i=1,\cdots,P,~\textrm{and}~x'_i=1\}$. 

{\it Example:} For $P=3$ and $x'_1x'_2x'_3=110$,
$\mathcal{S}_{110} = \mathcal{G}_1 \cup \mathcal{G}_2$, and $R(\mathcal{S}_{110})$ denotes the number of users reached by either $G_1$ or $G_2$.

\begin{definition}\label{def:column_z}
(Conditional independence)
Consider $P$ BGs and $\mathcal{G}_i$, the set of users reached by BG $G_i$, where $i=1,\cdots,P$. Suppose a user is reached by the BG $G_i$ with either the high probability $r_1(\mathcal{G}_i)=1-(1-r(\mathcal{G}_i))/d$ or the low probability $r_0(\mathcal{G}_i)=r(\mathcal{G}_i)/d$, where $d>1$ is a tuning parameter. Then for any user in the activity segment $\mathcal{U}_{x_1x_2\cdots x_P}$ defined in Definition \ref{def:segment}, the probability of that user reached by the subset $\mathcal{S}_{x'_1x'_2\cdots x'_P}$ defined in Definition \ref{def:minterm_maxterm} is given by:
\begin{eqnarray}\label{eqn:z_entry}
z(\mathcal{S}_{x'_1x'_2\cdots x'_P}, \mathcal{U}_{x_1x_2\cdots x_P})=1-\Pi_{\forall i,~x'_i=1}(1-r_{x_i}(\mathcal{G}_i)).
\end{eqnarray}
where $x'_1x'_2\cdots x'_P\in \mathcal{I}' \subseteq \mathcal{I} \setminus \{00\cdots 0\}$, 
$x_1x_2\cdots x_P\in \mathcal{I}$, and $\mathcal{I}$ is defined in Definition \ref{def:binary_strings}.
\end{definition}

{\it Remark:} (\ref{eqn:z_entry}) assumes that in each activity segment, the probability of a user being reached by one BG is independent of the others. In addition, we design the expressions of $r_1(\mathcal{G}_i)$ and $r_0(\mathcal{G}_i)$ to model the high-/low-reach probabilities for BG $G_i$ to reach any user in a segment. Such a binary assumption can be easily extended to create more activity segments. However, in this paper we consider this binary assumption only for the computation efficiency.

{\it Example:} Consider $P=3$, $x'_1x'_2x'_3=110$, and $x_1x_2x_3=100$. For any user in the segment $\mathcal{U}_{100}$, the BG $G_1$ has the high-reach probability $r_1(\mathcal{G}_1)$, and the BG $G_2$ has the low-reach probability $r_2(\mathcal{G}_1)$. Then the probability of the same user being reached by the subset $\mathcal{S}_{110}$ is given by $z(\mathcal{S}_{110}, \mathcal{U}_{100})=1-\Pi_{\forall i,~x'_i=1}(1-r_{x_i}(\mathcal{G}_i))=1-\Pi_{i=1}^2(1-r_{x_i}(\mathcal{G}_i))=1-(1-r_1(\mathcal{G}_1))(1-r_0(\mathcal{G}_2))$. 

With the above definitions, let us continue to describe our proposed model-based approach. Consider $P\geq 2$ BGs denoted by $G_i$,~where $i=1,\cdots,P$. Suppose we have measured the reach of $n$ union subsets $\mathcal{S}_{x'_1x'_2\cdots x'_P}$ where $x'_1x'_2\cdots x'_P \in \mathcal{I}' \subseteq \mathcal{I}\setminus \{00\cdots 0\}$, $n=|\mathcal{I}'|$, and $\mathcal{I}$ is defined in Definition \ref{def:binary_strings}. For brevity, we call these $n$ measurements the $n$ training points. Note that we require at least the reach of each single-BG and the all-BGs-union reach to be included, and thus $n\geq P+1$. That is, at least the $R(\mathcal{S}_{x'_1x'_2\cdots x'_P})$'s are available for all $x'_1x'_2\cdots x'_P$ satisfying $\sum_{i=1}^P x'_i=1$ (for the $P$ reach of single-BG) or $\sum_{i=1}^P x'_i = P$ (for the all-BGs-union reach). In addition, we may choose the reach of some other subsets, i.e., from the $R(\mathcal{S}_{x'_1x'_2\cdots x'_P})$'s where $1<\sum_{i=1}^P x'_i<P$, as additional training points. While we will discuss how to choose the additional training points in Sec. \ref{sec:framework}, we assume all the $n$ training points are given and fixed in this section. 

Our proposed model-based approach is introduced through the following algorithm.

\begin{algorithm}\label{alg:model_fit}
(The model-based approach)

Setup: Given the $n$ training points $R(\mathcal{S}_{x'_1\cdots x'_P})$'s for all $x'_1\cdots x'_P\in \mathcal{I}'$, calculate $r(\mathcal{S}_{x'_1\cdots x'_P})=R(\mathcal{S}_{x'_1\cdots x'_P})/U$. 
If the universe size $U$ is unknown, first estimate it with the proposed algorithm in (\ref{eqn:find_U}) in Sec. \ref{sec:framework}, and then treat the estimation $\hat{U}$ as $U$. Next, perform the following steps sequentially.
\begin{enumerate}
    \item Create $2^P+1$ mutually-exclusive activity segments according to Definition \ref{def:segment}. Since the last segment not reached by any BG does not play an active role, we ignore it in this algorithm.
    
    \item For each $x'_1\cdots x'_P\in \mathcal{I}'$ and each $x_1\cdots x_P\in \mathcal{I}$, calculate $z(\mathcal{S}_{x'_1\cdots x'_P}, \mathcal{U}_{x_1\cdots x_P})$ according to Definition \ref{def:column_z} 
    where the tuning parameter $d$ will be specified in Theorem \ref{theorem:zero_loss} and Sec. \ref{sec:framework}. After obtaining the $n\times 2^P$ entities, we form an $n\times 2^P$ matrix ${\bf Z}={\bf Z}(d)$ where each entry $z(\mathcal{S}_{x'_1\cdots x'_P},\mathcal{U}_{x_1\cdots x_P})$ is arranged in the row index in an ascending order w.r.t.  $\textrm{bin2dec}(x'_P\cdots x'_1)$ and the column index is $\textrm{bin2dec}(x_P\cdots x_1)$ (index starts from 0).
    
    \item Stack the $n$ entities $r(\mathcal{S}_{x'_1\cdots x'_P})$'s for all ${x'_1\cdots x'_P \in \mathcal{I}'}$ into an $n \times 1$ column vector ${\bf r}_{\mathcal{S}_{\mathcal{I}'}}$ by arranging the entries in an ascending order w.r.t. $\textrm{bin2dec}(x'_P\cdots x'_1)$. 
    
    \item Denote by ${\bf w}$ a $2^P\times 1$ column vector and let $J_d({\bf w})\triangleq \|{\bf r}_{\mathcal{S}_{\mathcal{I}'}}-{\bf Z}(d)\cdot {\bf w}\|_2^2$ be the training error. Then solve the following non-negative least squares (NNLS) problem: 
    \begin{eqnarray}\label{eqn:NNLS}
    {\bf w}^*= &\!\!\!\!\!\!\!\!& \arg \min_{\bf w} J_d({\bf w}),\\
    &\!\!\!\!\!\!\!\!& \textrm{s.t.} ~~~~\left\{\begin{array}{r} {\bf w} \succeq {\bf 0},\\
    \|{\bf w}\|_1\leq 1.\end{array}\right.\notag
    \end{eqnarray}
    Here, ${\bf w}^*$, which our proposed model is trained for, is the weight vector to linearly combine all the reach probabilities in the $2^P$ activity segments. 
    
    \item To estimate the reach of any unobserved subset $\mathcal{S}_q$ where $q \in \mathcal{I}\setminus \mathcal{I}'$, calculate $z(\mathcal{S}_q,\mathcal{U}_{x_1\cdots x_P})$ for each $x_1\cdots x_P\in \mathcal{I}$ following Definition \ref{def:column_z}. Then form a column vector ${\bf z}(q)$ by placing $z(\mathcal{S}_q,\mathcal{U}_{x_1\cdots x_P})$ at the $(\textrm{bin2dec}(x_P\cdots x_1))^{th}$ entry (index starts from 0). Finally, the reach estimation of the subset $\mathcal{S}_q$ is given by:
    \begin{eqnarray}
    \hat{r}(\mathcal{S}_q)= {\bf z}^T(q)\cdot {\bf w}^*\Longrightarrow \hat{R}(\mathcal{S}_q)= \hat{r}(\mathcal{S}_q)\cdot U={\bf z}^T(q)\cdot {\bf w}^*\cdot U.
    \end{eqnarray}
\end{enumerate}
\end{algorithm}

{\it Remark:} In Step 1 above, because the universe is divided into $2^P+1$ mutually-exclusive activity segments, the weights used for linearly combining the reach probabilities in each segment should sum to one. Since we exclude the non-active segment after Step 1, the constraint $\|{\bf w}\|_1\leq 1$ is imposed in Step 4. In addition, the functions $r_1(\mathcal{G}_i)$ and $r_0(\mathcal{G}_i)$ in Step 2 and Definition \ref{def:column_z} are used as a simple way to model the high- and low-probability for BG $G_i$ to reach a user, and the values can be tuned via the model parameter $d$.

\begin{theorem}\label{theorem:zero_loss} 
(Perfect model-fitting)
Suppose the $n$ training points in Algorithm \ref{alg:model_fit} are consistent. For the problem formulated in (\ref{eqn:NNLS}) in Algorithm \ref{alg:model_fit}, if $d$ is large enough, there always exists a solution ${\bf w}^*$ so that $J_d({\bf w}^*)=0$.
\end{theorem}

{\it Proof:} The proof is deferred to Appendix A. 

{\it Remark:} Theorem \ref{theorem:zero_loss} implies if $d$ is large enough, we can perfectly fit the $n$ training points to the model, i.e., ${\bf r}_{\mathcal{S}_{\mathcal{I}'}}={\bf Z}(d)\cdot {\bf w}^*$. This is accomplished by representing the observational vector ${\bf r}_{\mathcal{S}_{\mathcal{I}'}}$ with a linear combination of the columns of the matrix ${\bf Z}(d)$, and the combination weights are ${\bf w}^*$. Recall that  ${\bf Z}(d)$ has $2^P$ columns, each corresponding to one activity segment created in Step 1. We can also interpret Theorem \ref{theorem:zero_loss} from the geometric perspective. In the $n$-dimensional vector space, the $2^P$ columns of ${\bf Z}(d)$, as $2^P$ vertices, define a convex hull. Then the $n$-dimensional point represented by ${\bf r}_{\mathcal{S}_{\mathcal{I}'}}$ must stay in this convex hull. To help the reader interpret Algorithm \ref{alg:model_fit} and Theorem \ref{theorem:zero_loss}, we provide two examples in Sec. \ref{sec:theorem_examples}.

Be aware that Theorem \ref{theorem:zero_loss} requires the assumption of the reach consistency among the training points to achieve zero training error, but Algorithm \ref{alg:model_fit} does not need it. In fact, even if the training points are inconsistent, Algorithm \ref{alg:model_fit} can still be applied, but the training error will no longer be zero no matter how large $d$ is. In practice, if the training points are inconsistency due to, e.g., measurement error and/or the differential-privacy (DP) noise, we may need extra processing to adjust the training points and/or the reach estimations of any unobserved subset when necessary. 


\subsection{Understanding Algorithm \ref{alg:model_fit} and Theorem \ref{theorem:zero_loss}}\label{sec:theorem_examples}

To better interpret our proposed model-based approach, we provide two examples with $P=2$ and $P=3$ in this section, followed by the insights of Algorithm \ref{alg:model_fit}. The example with $P=2$ focuses on how to fit the data to our model. It is the simplest example that can be easily visualized in a 3-D system. In the example with $P=3$, we demonstrate the algorithm in terms of model fitting and reach estimation. 

\subsubsection{Example 1: $P=2$ BGs} \label{sec:example_algorithm_fitting}

For $P=2$ there are three subsets $\mathcal{S}_{x'_1x'_2}$ for all $x'_1x'_2\in \mathcal{I}'=\{10,01,11\}$ by ignoring the subset of users not reached by any BG. Since the reach of all of them have to be used as the training points, there is no more subset for reach estimation. Thus, we only show the algorithm flow until Step 4 in the following.

\begin{enumerate}
    \item For $P=2$ there are a total of $2^P=2^2=4$ activity segments, denoted by $\mathcal{U}_{x_1x_2}$ for all $x_1x_2\in \mathcal{I}=\{00,10,01,11\}$.
    
    \item As described in Sec. \ref{sec:improved_DMM}, the must-be-included three training points are the $R(\mathcal{S}_{x'_1x'_2})$'s for all $x'_1x'_2\in \mathcal{I}'=\{10,01,11\}$. According to Definition \ref{def:minterm_maxterm}, we specify $\mathcal{S}_{10}=\mathcal{G}_1$, $\mathcal{S}_{01}=\mathcal{G}_2$, and $\mathcal{S}_{11}=\mathcal{G}_1\cup \mathcal{G}_2$. With Definition \ref{def:column_z}, ${\bf z}(\mathcal{S}_{x'_1x'_2}, \mathcal{U}_{x_1x_2})$ for $x'_1x'_2=10,01,11$ and $x_1x_2=00,10,01,11$ can be calculated as follows:
    \begin{itemize}
        \item $x'_1x'_2=10$, $\mathcal{S}_{x'_1x'_2}=\mathcal{G}_1\Longrightarrow$ $z(\mathcal{S}_{10},\mathcal{U}_{00})=z(\mathcal{S}_{10},\mathcal{U}_{01})=r_0(\mathcal{G}_1)=r(\mathcal{G}_1)/d$,   $z(\mathcal{S}_{10},\mathcal{U}_{10})=z(\mathcal{S}_{10},\mathcal{U}_{11})=r_1(\mathcal{G}_1)=1-(1-r(\mathcal{G}_1))/d$. 
    
        \item $x'_1x'_2=01$,  $\mathcal{S}_{x'_1x'_2}=\mathcal{G}_2\Longrightarrow$ $z(\mathcal{S}_{01},\mathcal{R}_{00})=z(\mathcal{S}_{01},\mathcal{R}_{10})=r(\mathcal{G}_2)/d$ and also $z(\mathcal{S}_{01},\mathcal{R}_{01})=z(\mathcal{S}_{01},\mathcal{R}_{11})=1-(1-r(\mathcal{G}_2))/d$.
    
        \item $x'_1x'_2=11$, $\mathcal{S}_{x'_1x'_2}=\mathcal{G}_1\cup \mathcal{G}_2\Longrightarrow$ $z(\mathcal{S}_{11},\mathcal{R}_{x_1x_2})=1-(1-z(\mathcal{S}_{10},\mathcal{R}_{x_1x_2}))(1-z(\mathcal{S}_{01},\mathcal{R}_{x_1x_2}))$ for $x_1x_2=00,10,01,11$. 
    \end{itemize}
    
    Then according to the row order $x'_1x'_2=10,01,11$ and the column order $x_1x_2=00,10,01,11$ sequentially, we form the ${\bf Z}(d)$ matrix as:
    \begin{eqnarray}\label{eqn:P2_Zmat_general}
    {\bf Z}(d)=\left[\begin{array}{cccc} \frac{r_1}{d}&1-\frac{1-r_1}{d}&\frac{r_1}{d}&1-\frac{1-r_1}{d}\\
    \frac{r_2}{d}&\frac{r_2}{d}&1-\frac{1-r_2}{d}&1-\frac{1-r_2}{d}\\
    1\!-\!(1\!-\!\frac{r_1}{d})(1\!-\!\frac{r_2}{d})&1\!-\!\frac{1-r_1}{d}(1\!-\!\frac{r_2}{d})&1\!-\!(1\!-\!\frac{r_1}{d})\frac{1-r_2}{d}&1\!-\!\frac{(1-r_1)(1-r_2)}{d^2}
    \end{array}\right].
    \end{eqnarray}

    \item Form the vector ${\bf r}_{\mathcal{S}_{\mathcal{I}'}}=[r(\mathcal{S}_{10}), r(\mathcal{S}_{01}), r(\mathcal{S}_{11})]^T=[R(\mathcal{S}_{10})/U, R(\mathcal{S}_{01})/U, R(\mathcal{S}_{11})/U]^T$ from the three training points.
    
    \item Choose a large enough $d$, say, $d\rightarrow +\infty$ as the extreme case. Then ${\bf Z}(d)$ in Step 2 becomes:
    \begin{eqnarray}\label{eqn:P2_Zmat}
    {\bf Z}(+\infty)=\left[\begin{array}{cccc} 
    0 & 1 & 0 & 1 \\
    0 & 0 & 1 & 1 \\ 
    0 & 1 & 1 & 1 
    \end{array}\right].
    \end{eqnarray}
    With the above ${\bf Z}(+\infty)$, solve the NNLS problem formulated in (\ref{eqn:NNLS}). Denote the solution by ${\bf w}^*$, and it can be verified that $J_d({\bf w}^*)=0$. 
\end{enumerate}

From (\ref{eqn:P2_Zmat_general}) to (\ref{eqn:P2_Zmat}), the first column of ${\bf Z}(d)$ becomes all zero when $d\rightarrow +\infty$, which means both BGs in the segment $\mathcal{U}_{00}$ have zero low-reach probability. Thus, $\mathcal{U}_{00}$ merges with the non-active segment that we introduced but excluded after Step 1 of Algorithm \ref{alg:model_fit}. Also, only in this case, each activity segment $\mathcal{U}_{x_1x_2}$ is equivalent to the primitive region $\mathcal{R}_{x_1x_2}$ defined in Definition \ref{def:pr}.

\begin{figure}[!t]
\centering\vspace{-0.1in}
\includegraphics[scale=0.35]{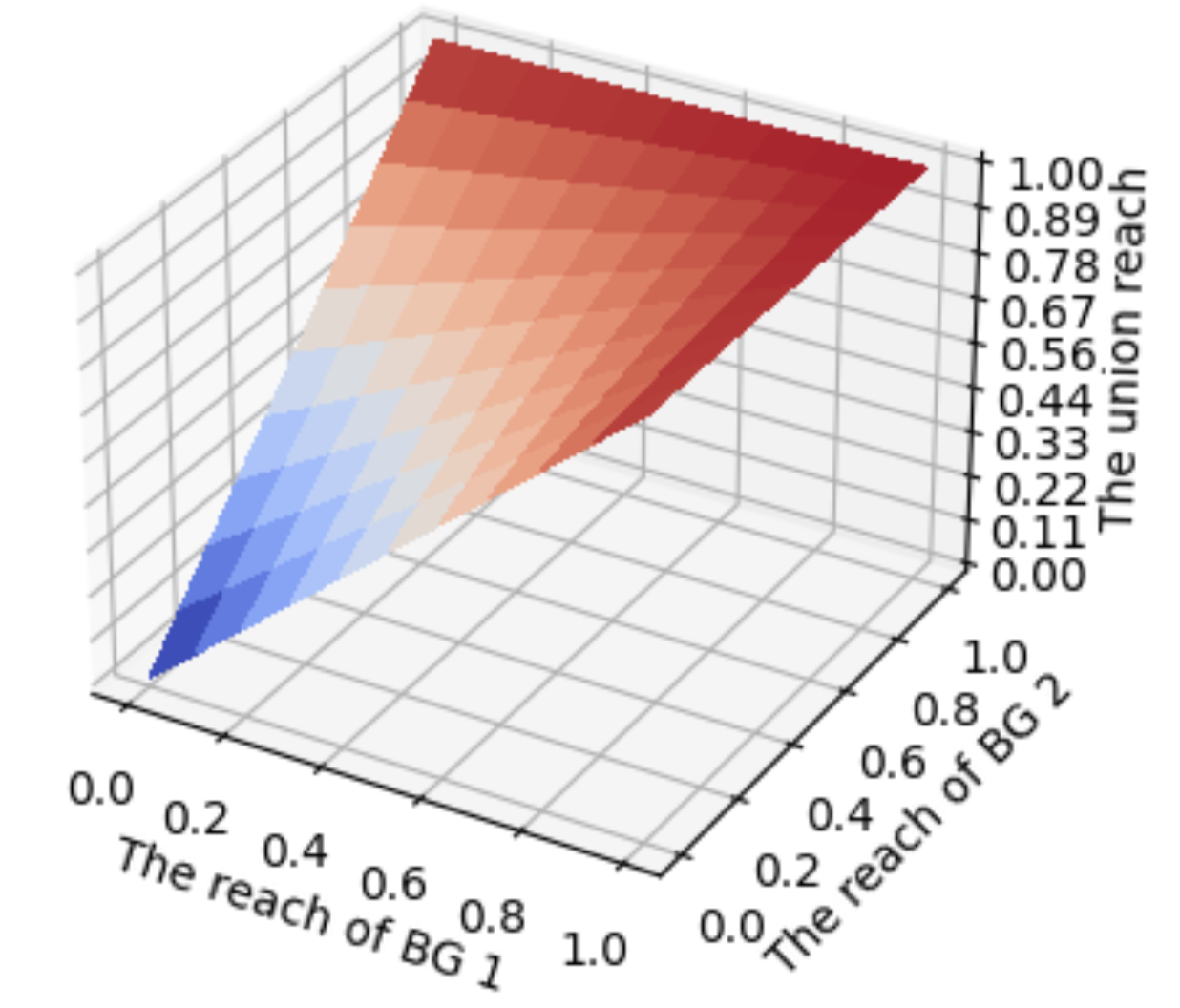}
\vspace{-0.1in}
\caption{Visualization of the reach surface for the proposed model with $P=2$}
\label{fig:perfect_fit_P2}
\end{figure}

In Figure \ref{fig:perfect_fit_P2}, we show the four points $(0,0,0)$, $(1,0,1)$, $(0,1,1)$, $(1,1,1)$ by treating each column of ${\bf Z}(+\infty)$ as the three coordinates in the 3-D system. Each point comes from one out of the four activity segments that we created in Step 1. These four points, as four vertices, define a 3-D convex hull. The convex hull is enclosed by four surfaces, and each surface is established by every three of the four vertices above. As long as the training points are consistent, the observed vector ${\bf r}_{\mathcal{S}_{\mathcal{I}'}}$ must stay in the formed convex hull. In fact, each of the established four surfaces produces an underlying triangle inequality associated with the reach of the three subsets. 

Additionally, the colored surface between the four vertices in Figure \ref{fig:perfect_fit_P2} is obtained from (\ref{eqn:z_entry}), which is given by $z=1-(1-x)(1-y)$ where $x$, $y$, $z$ represent the reach of BG1, reach of BG2, and their union reach. 
In particular, by letting $(x,y)=(0,0),~(1,0),~(0,1),~(1,1)$, we obtain the four vertices above sequentially; and if $x$ and $y$ take fractions between zero and one, we obtain the colored surface shown in Figure \ref{fig:perfect_fit_P2}. Clearly, the proposed modeling approach relies on the assumption that the probabilities for the two BGs to reach a user are independent in each activity segment. Hence, any point on the colored surface also satisfies the reach consistency, and thus the colored surface also stays in the formed convex hull. 

Note that the colored surface obtained from our modeling approach does not necessarily mean the ground truth. Also, the four vertices represent the extreme cases. Back to Theorem \ref{theorem:zero_loss}, the condition of large enough $d$ implies that we can find a finite $d$ to achieve perfect model fitting. With the geometric interpretation of Figure \ref{fig:perfect_fit_P2}, it means when $d$ becomes smaller, the four vertices will be away from the extreme points and move along the colored surface. As a result, the volume of the convex hull becomes smaller, but zero training error can still be achieved as long as the shrinking convex hull stills include the vector ${\bf r}_{\mathcal{S}_{\mathcal{I}'}}$.

\subsubsection{Example 2: $P=3$ BGs}

Suppose we have $n=4$ training points $R(\mathcal{S}_{x'_1x'_2x'_3})$'s for all $x'_1x'_2x'_3\in \mathcal{I}'=\{100,010,001,111\}$, and we can obtain the corresponding $r(\mathcal{S}_{x'_1x'_2x'_3})=R(\mathcal{S}_{x'_1x'_2x'_3})/U$. Also, suppose $R(\mathcal{S}_{110})$ is unobserved and we want to estimate it. Then we show each step of Algorithm \ref{alg:model_fit} as follows.

\begin{enumerate}
    \item For $P=3$ there are a total of $2^P=2^3=8$ activity segments, denoted by $\mathcal{U}_{x_1x_2x_3}$ for all $x_1x_2x_3\in \mathcal{I}=\{000,100,010,110,001,101,011,111\}$.
    
    \item Considering $\mathcal{I}'$ from the four training points. Following Definition \ref{def:minterm_maxterm} we readily specify $\mathcal{S}_{100}=\mathcal{G}_1$, $\mathcal{S}_{010}=\mathcal{G}_2$, $\mathcal{S}_{001}=\mathcal{G}_3$, and $\mathcal{S}_{111}=\cup_{i=1}^3 \mathcal{G}_i$. With Definition \ref{def:column_z}, ${\bf z}(\mathcal{S}_{x'_1x'_2x'_3}, \mathcal{U}_{x_1x_2x_3})$ for each $x'_1x'_2x'_3\in \mathcal{I}'$ and each $x_1x_2x_3\in \mathcal{I}$ can be calculated as follows:
    \begin{itemize}
        \item $x'_1x'_2x'_3=100$, $\mathcal{S}_{x'_1x'_2x'_3}=\mathcal{G}_1\Longrightarrow$ $z(\mathcal{S}_{100},\mathcal{U}_{x_1x_2x_3})=r(\mathcal{G}_1)/d$ for $x_1x_2x_3=000$, $010$, $001$, $011$, and $z(\mathcal{S}_{100},\mathcal{U}_{x_1x_2x_3})=1-(1-r(\mathcal{G}_1))/d$ for $x_1x_2x_3=100, 110, 101, 111$. 
    
        \item $x'_1x'_2x'_3=010$, $\mathcal{S}_{x'_1x'_2x'_3}=\mathcal{G}_2\Longrightarrow$
        $z(\mathcal{S}_{010},\mathcal{U}_{x_1x_2x_3})=r(\mathcal{G}_2)/d$ for $x_1x_2x_3=000$, $100$, $001$, $101$, and $z(\mathcal{S}_{010},\mathcal{U}_{x_1x_2x_3})=1-(1-r(\mathcal{G}_2))/d$ for $x_1x_2x_3=010, 110, 011, 111$. 
    
        \item $x'_1x'_2x'_3=001$, $\mathcal{S}_{x'_1x'_2x'_3}=\mathcal{G}_3\Longrightarrow$
        $z(\mathcal{S}_{001},\mathcal{U}_{x_1x_2x_3})=r(\mathcal{G}_3)/d$ for $x_1x_2x_3=000$, $100$, $010$, $110$, and $z(\mathcal{S}_{001},\mathcal{U}_{x_1x_2x_3})=1-(1-r(\mathcal{G}_3))/d$ for $x_1x_2x_3=001, 101, 011, 111$. 
    
        \item $x'_1x'_2x'_3=111$, $\mathcal{S}_{x'_1x'_2x'_3}=\cup_{i=1}^3\mathcal{G}_i\Longrightarrow$ $z(\mathcal{S}_{111},\mathcal{U}_{x_1x_2x_3})=1-\Pi_{x'_1x'_2x'_3\in \{100,010,001\}}(1-z(\mathcal{S}_{x'_1x'_2x'_3},\mathcal{U}_{x_1x_2x_3}))$ for each $x_1x_2x_3\in \mathcal{I}$. 
    \end{itemize}
    Then according to the row order $x'_1x'_2x'_3=100,010,011,111$ and the column order $x_1x_2x_3=000,100,010,110,001,101,011,111$, we form the ${\bf Z}(d)$ matrix as:
    \begin{eqnarray}\label{eqn:P3_Zmat}
    {\bf Z}(d)=\left[\begin{array}{cccc} z(\mathcal{S}_{100},\mathcal{U}_{000})&z(\mathcal{S}_{100},\mathcal{U}_{100})&\cdots &z(\mathcal{S}_{100},\mathcal{U}_{111})\\
    z(\mathcal{S}_{010},\mathcal{U}_{000})&z(\mathcal{S}_{010},\mathcal{U}_{100})&\cdots &z(\mathcal{S}_{010},\mathcal{U}_{111})\\
    z(\mathcal{S}_{001},\mathcal{U}_{000})&z(\mathcal{S}_{001},\mathcal{U}_{100})&\cdots &z(\mathcal{S}_{001},\mathcal{U}_{111})\\
    z(\mathcal{S}_{111},\mathcal{U}_{000})&z(\mathcal{S}_{111},\mathcal{U}_{100})&\cdots &z(\mathcal{S}_{111},\mathcal{U}_{111})
    \end{array}\right].
    \end{eqnarray}
    
    \item Form the vector ${\bf r}_{\mathcal{S}_{\mathcal{I}'}}=[r(\mathcal{S}_{100}), r(\mathcal{S}_{010}), r(\mathcal{S}_{001}), r(\mathcal{S}_{111})]^T$ from the training points.

    \item For any given $d$, solve the NNLS problem formulated in (\ref{eqn:NNLS}), and denote the solution by ${\bf w}^*$. Theorem \ref{theorem:zero_loss} implies  $J_d({\bf w}^*)=0$ if $d$ is large enough. We can directly let $d\rightarrow +\infty$ like in the example with $P=2$ or to find the minimum $d$ so that $J_d({\bf w}^*)=0$. 

    \item For the unobserved subset $\mathcal{S}_{110}=\mathcal{G}_1\cup \mathcal{G}_2\setminus \mathcal{G}_3$, we calculate $z(\mathcal{S}_{110},\mathcal{U}_{x_1x_2x_3})=1-(1-z(\mathcal{S}_{100},\mathcal{U}_{x_1x_2x_3}))(1-z(\mathcal{S}_{010},\mathcal{R}_{x_1x_2x_3}))$ for each $x_1x_2x_3\in \mathcal{I}$. Then arrange them into a column vector ${\bf z}(110)=[z(\mathcal{S}_{110},\mathcal{U}_{000}), z(\mathcal{S}_{110},\mathcal{U}_{100}),\cdots, z(\mathcal{S}_{110},\mathcal{U}_{111})]^T$ following the order of $\mathcal{U}_{x_1x_2x_3}$ with $x_1x_2x_3=000,100,010,110,001,101,011,111$. Finally, the reach of the subset $\mathcal{S}_{110}$ is estimated as $\hat{R}(\mathcal{S}_{110})=\hat{r}(\mathcal{S}_{110})\cdot U = {\bf z}^T(110)\cdot {\bf w}^*\cdot U$.
\end{enumerate}

{\it Remark:} In Step 2 above, the independence-model assumption is applied in each of the eight activity segments. That is,  $h=1-(1-x)(1-y)(1-z)$ where $x$, $y$, $z$, $h$ represent the reach of BG1, that of BG2, that of BG3, and their union reach, in each segment. Enumerating $x$, $y$, $z$ over their high- and low-reach probabilities, we obtain a total of 8 points $(x,y,z,h)$'s in the $4$-dimensional vector space. These 8 points, as 8 vertices, define a convex hull in the 4-D system. Theorem \ref{theorem:zero_loss} implies that the vector ${\bf r}_{\mathcal{I}'}$ must stay in this convex hull, if the training points are consistent.

\subsubsection{Why do we Need Algorithm \ref{alg:model_fit}?}



Without the universe segmentation, the conditional independence model reduces to the independence model, which is simpler and widely used in practice. Compared to the independence model, the conditional independence model provides higher flexibility than the independence model in characterizing the correlation between multiple BGs. However, in practice, without any prior correlation information between the BGs, it is usually difficult to determine the number of segments needed, the weights of each segment, and the single-BG reach in each segment.

In Algorithm \ref{alg:model_fit}, we utilize the conditional independence model and address the above challenges through a modeling approach. Specifically, we divide the universe into $2^P$ mutually-exclusive activity segments by assuming each BG has low/high-reach probability in a segment and a non-active segment for the users not reached by any BG. In each segment labeled with $x_1\cdots x_P$, the single-BG reach probability is modeled as $r_{x_i}(\mathcal{G}_i)$ defined in Definition \ref{def:column_z}. Then the probability of a user being reached by a subset is computed from the modeled single-BG reach probabilities in  (\ref{eqn:z_entry}) with the independence-model assumption. Finally, the combination weights of each segment are obtained through solving the NNLS problem formulated in (\ref{eqn:NNLS}).

\section{Integration of the Two Approaches for Subset-Reach Estimation}\label{sec:framework}


In Sec. \ref{sec:model_free} and Sec. \ref{sec:DMM}, we introduce a model-free and a model-based approaches for estimating the reach of a subset. While the former provides a $100\%$-confidence interval as a coarse estimate, the latter provides a point estimate. In practice, whenever the advertiser asks for the reach of a subset, we can first use the former to tell an estimation range as narrower as possible, and then use the latter to provide answer without uncertainty. With this idea, in this section, we integrate the model-free and model-based approaches into a framework to estimate the reach of any subset. As shown in Figure \ref{fig:framework}, the proposed framework consists of six blocks, each with the following functionality:
\begin{enumerate}
    \item ``Linear programming solver": It has been introduced in Theorem \ref{theorem:bound_subset_reach} in Sec. \ref{sec:model_free}.
    
    \item ``Adaptively choose training points": What if we are allowed to choose some training points? We propose an algorithm by applying the linear programming solver to sequentially choose the next training point as the one with the largest uncertainty. See Sec. \ref{sec:adap_select_training} in detail.
    
    \item ``Conditional independence model": It has been introduced in Sec. \ref{sec:improved_DMM}, particularly in Steps 1--4 of Algorithm \ref{alg:model_fit}. If the universe size $U$ is unavailable, we propose a very simple algorithm to estimate it. See Sec. \ref{sec:est_universe_size} in detail.
    
    \item ``Tune model parameters via cross validation": What is a proper value of the model parameter $d$ we should use in Algorithm \ref{alg:model_fit}? We propose an algorithm to answer this question. The idea is to leverage cross validation and choose the $d$ that leads to the minimum average cross-validation error. See Sec. \ref{sec:tuning_d} in detail.
    
    \item ``Model estimation": We apply both the model-free approach in Theorem \ref{theorem:bound_subset_reach} and the model-based approach in Step 5 of Algorithm \ref{alg:model_fit} to estimate the reach of any unobserved subset. The estimations include a $100\%$-confidence interval and a point estimate.
    
    \item ``Error bar via cross validation": How to evaluate the model-estimation quality? We answer this question by proposing an algorithm to define an error-bar metric. The idea is to analyze the estimation accuracy on the testing points based on the statistics of the cross-validation accuracy on the training points. See Sec. \ref{sec:error_bar} in detail.
\end{enumerate} 


Before proceeding into the description of the four algorithms, we first summarize the additional notations used throughout this section in Table \ref{tab:framework_notations}.
\begin{table}[H]
\centering
\caption{Additional notations used throughout this section}
\vspace{0.5em}
\begin{tabular}{p{5em}p{35em}}
    \hline\hline
    Notations & Description \\ \hline
    $\mathcal{I}$ & the binary string set $\{x_1\cdots x_P~|~x_i\in\{0,1\},i=1,\cdots,P\}$ defined in Definition \ref{def:binary_strings}\\ 
    $\mathcal{I}'_b$ & the basic binary string set  $\{x'_1\cdots x'_P~|~\sum_{i=1}^P x'_i=1~ \textrm{or}~ \sum_{i=1}^P x'_i=P\}$ used to define the must-be-included training points, i.e., $|\mathcal{I}'_b|=P+1$\\ 
    $\mathcal{I}_c$ & the candidate string set defined as $\mathcal{I}\setminus \{00\cdots 0\} \setminus \mathcal{I}'_b$ for training points selection\\
    $\mathcal{I}'$ & the binary string set used to define the $n$ training points, i.e., $|\mathcal{I}'|=n$ and $\mathcal{I}'_b\subseteq \mathcal{I}'\subseteq \mathcal{I}\setminus \{00\cdots 0\}$\\ 
    $\overline{R}(\mathcal{S}_q), \underline{R}(\mathcal{S}_q)$ & the upper and the lower bounds of $R(\mathcal{S}_q)$ obtained from Theorem \ref{theorem:bound_subset_reach} \\ 
    \hline\hline
\end{tabular}\label{tab:framework_notations}
\end{table}

Consider the following example with $P=3$ and $n=5$ to interpret the notations in Table \ref{tab:framework_notations}:
\begin{eqnarray*}
\mathcal{I} &\!\!\!\!=\!\!\!\!& \{100,010,110,001,101,011,111,000\},\\
\mathcal{I}'_b &\!\!\!\!=\!\!\!\!& \{100,010,001,111\},\\
\mathcal{I}_c &\!\!\!\!=\!\!\!\!& \{110,101,011\},\\
\mathcal{I}' &\!\!\!\!=\!\!\!\!& \{100,010,001,111,101\}.
\end{eqnarray*}
Regarding the $n=5$ training points, at least, the reach of the four subset $R(\mathcal{S}_{x'_1x'_2x'_3})$'s for all $x'_1x'_2x'_3\in \mathcal{I}'_b$ must be included. Thus, we are allowed to choose the reach of one more subset $R(\mathcal{S}_{x'_1x'_2x'_3})$ as an additional training point by choosing one $x'_1x'_2x'_3\in \mathcal{I}_c$. For example, if we choose $x'_1x'_2x'_3=101$, then the $n=5$ training points are given by $R(\mathcal{S}_{x'_1x'_2x'_3})$'s for all $x'_1x'_2x'_3\in \mathcal{I}'$. Moreover, recall that $\mathcal{I}$ firstly introduced in Definition \ref{def:binary_strings} is used to define the universe segmentation in Definition \ref{def:segment} and also to define a subset $\mathcal{S}_{x'_1\cdots x'_P}$ in Definition \ref{def:minterm_maxterm}.

\subsection{Adaptively Choose Training Points}\label{sec:adap_select_training}

The training points, as introduced in Sec. \ref{sec:improved_DMM}, must contain all the $P$ single-BG reach and the all-BGs-union reach. Besides these $P+1$ training points, when the privacy budget allows, one can measure the reach of more subsets and use them as the extra training points to improve the model accuracy. Nevertheless, how to determine the additional training points? 
In this section, we propose an algorithm to heuristically answer this question. The main idea is to leverage the linear programming solver introduced in Theorem \ref{theorem:bound_subset_reach} in Sec. \ref{sec:model_free}.  

We begin with the problem formulation. Consider $P$ BGs, denoted by $G_i,~i=1,\cdots,P$, and the users reached by $G_i$ is denoted by $\mathcal{G}_i$. If the total number of the training points $n=P+1$, we cannot choose any more training points. Otherwise, we desire to measure the reach of additional $n-(P+1)$ subsets $R(\mathcal{S}_{x'_1\cdots x'_P})$'s by choosing $n-(P+1)$ different indices $x'_1\cdots x'_P$'s from the candidate string index set $\mathcal{I}_c$. To this end, we propose the following algorithm:

\begin{algorithm} \label{alg:adaptive_select_train}
(Adaptively choosing the training points)
\begin{enumerate}
    \item Initially, let $\mathcal{I}'=\mathcal{I}_b'$ as a start, $\mathcal{T}\triangleq \{R(\mathcal{S}_{q'})|q'\in\mathcal{I}'_b\}$ be the $P+1$ training points we must have, and $\mathcal{I}_c= \mathcal{I}\setminus \{00\cdots 0\} \setminus \mathcal{I}'_b$ be the candidate-index set. 
    
    \item For each $q\in \mathcal{I}_c$, based on $\mathcal{T}$, calculate $\overline{R}(\mathcal{S}_q)$ and  $\underline{R}(\mathcal{S}_q)$ based on the model-free approach in Theorem \ref{theorem:bound_subset_reach}.
    Choose the one with the largest bound gap as the next training point, i.e., 
    \begin{eqnarray}
    q^*=\arg\max_{q\in \mathcal{I}_c}|\overline{R}(\mathcal{S}_q)-\underline{R}(\mathcal{S}_q)|.
    \end{eqnarray}  
    
    \item Measure $R(\mathcal{S}_{q^*})$. Then update $\mathcal{I}'=\mathcal{I}'\cup \{q^*\}$,  $\mathcal{T}=\mathcal{T}\cup \{R(\mathcal{S}_{q^*})\}$, and $\mathcal{I}_c=\mathcal{I}_c\setminus \{q^*\}$. 
    
    \item If $|\mathcal{I}'|=n$, the algorithm terminates; otherwise, go back to Step 2. 
\end{enumerate}
\end{algorithm}

{\it Remark:} In Step 2 of Algorithm \ref{alg:adaptive_select_train}, the larger gap between the bounds, the higher uncertainty that reach estimation. To reduce the estimation uncertainty as much as possible, we choose the subset that produces the largest bound gap, measure the reach of that subset, and then add it to the training points. As a result, Algorithm \ref{alg:model_fit} can produce a narrower $100\%$-confidence interval when estimating the reach of any unobserved subset.

\subsection{Estimating the Universe Size}\label{sec:est_universe_size}

In Algorithm \ref{alg:model_fit}, it still requires to specify the universe size $U$, if it is not available in practice. Without any prior information on $U$, we propose the following algorithm to estimate $U$.
\begin{algorithm}
(Estimate the universe size)

The universe size is estimated by solving the following equation w.r.t. $U$:
\begin{eqnarray}\label{eqn:find_U}
1-\frac{R(\cup_{i=1}^P \mathcal{G}_i)}{U}=\Pi_{i=1}^P\left(1-\frac{R(\mathcal{G}_i)}{U}\right).
\end{eqnarray}
\end{algorithm}

{\it Remark:} (\ref{eqn:find_U}) requires all the $P$ single-BG reach and the all-BGs-union reach. Without any prior information on $U$, we choose $U$ to satisfy the independence-model assumption. For $P\leq 5$, (\ref{eqn:find_U}) can be solved with a close-formed expression; otherwise, it can be resolved via optimization algorithms, such as binary search or the Newton–Raphson method.

\subsection{Tuning the Parameter $d$ Used in Algorithm \ref{alg:model_fit} via Cross Validation}\label{sec:tuning_d}

In Theorem \ref{theorem:zero_loss}, we show that the training error can be zero if $d$ is large enough. However, it does not indicate how the model will perform on the testing points/dataset. Note that we ultimately want to estimate the reach of any unobserved subset as accurately as possible. Thus, the model generalization capability is critical. If $d$ is large enough so that the training error is zero, when the model is applied to estimate the reach of any unobserved subset, over-fitting tends to happen. On the other hand, if $d$ is too small so that the training error is significantly larger than zero, then under-fitting is likely to happen. Considering the trade-off between over-fitting and under-fitting, how to choose a \emph{proper} value for $d$? 

If the number of the training points is very large, we can split the training points into two mutually-exclusive subsets, one for training and the other for validation. Training the model on the training subset and fine-tuning model parameters on the validation subset will produce a more reliable solution. Nevertheless, we often only have a small number of the training points. To better exploit the limited training points, we borrow the idea of $K$-fold cross validation. Specifically, out of the $n$ training points, excluding the $P$+1 must-be-used training points, the rest $n-(P+1)$ training points, i.e., the $R(\mathcal{S}_{x'_1\cdots x'_P})$'s for all $x'_1\cdots x'_P\in \mathcal{I}'\setminus \mathcal{I}'_b$ can be used for cross validation. In each round of cross validation, we use one of them as the validation point and the other $n-1$ training points to train the model following Algorithm \ref{alg:model_fit}. 

During the training process in each round of cross validation, we can perfectly fit all the $n-1$ training points to the model to achieve zero training error by searching for the minimum $d$ value. Such a way requires iterative computation and still inherits the potential over-fitting problem. To make the optimization simpler, an alternatively way is to train the model with a fixed set of pre-defined $d$ values, and then the optimal $d$ will be the one which produces the minimum averaged estimation error over all the validation rounds. With the above idea, we propose the following algorithm to determine the value of $d$ used in Algorithm \ref{alg:model_fit}.

\begin{algorithm}\label{alg:k_fold_val}
(Tuning $d$ via cross validation)

Setup: Define the minimum, the maximum, and the total number of the $d$ values we consider, denoted by $d_{\min}$, $d_{\max}$, and $D$, respectively. Then the set of $d$ values that we consider is given by
\begin{eqnarray}
\mathcal{D}\triangleq \left\{d_{\min}+c\cdot \frac{d_{\max}-d_{\min}}{D-1}~\big|~c=0,\cdots,D-1\right\}.
\end{eqnarray}
In this paper, we assume $d_{\min}=1$, $d_{\max}=5$, and $D=10$. Then for each $d\in \mathcal{D}$, we perform a total of $|\mathcal{I}'\setminus \mathcal{I}'_b|=n-(P+1)$ rounds of cross validation. In each round,
\begin{enumerate}
    \item Choose one $q'\in \mathcal{I}'\setminus \mathcal{I}'_b$ and use $R(\mathcal{S}_{q'})$ as the ground truth for validation.
    
    \item Fit the other $n-1$ training points $R(\mathcal{S}_{x'_1\cdots x'_P})$'s for all $x'_1\cdots x'_P\in \mathcal{I}'\setminus \{q'\}$ to the model following Steps 1--4 of Algorithm \ref{alg:model_fit}, and estimate $R(\mathcal{S}_{q'})$ following Step 5 of Algorithm \ref{alg:model_fit}. Denote the estimation by $\hat{R}_d(\mathcal{S}_{q'})$, where the footnote of $d$ indicates this estimation is $d$-dependent. 
    \item Use Theorem \ref{theorem:bound_subset_reach} to obtain $\overline{R}(\mathcal{S}_{q'})$ and $\underline{R}(\mathcal{S}_{q'})$.
    
    \item Compute the relative error of the reach estimation $\hat{R}_d(\mathcal{S}_{q'})$ as
    \begin{eqnarray}\label{eqn:rel_err}
    e_{\textrm{rel}}(\hat{R}_d(\mathcal{S}_{q'})) \triangleq \frac{\hat{R}_d(\mathcal{S}_{q'})-R(\mathcal{S}_{q'})}{\overline{R}(\mathcal{S}_{q'})-\underline{R}(\mathcal{S}_{q'})}.
    \end{eqnarray}
\end{enumerate}
After completing all the $n-(P+1)$ rounds of validation, we calculate the average of the absolute validation relative error. Then the optimal $d^*$ is selected as one that produces the minimum averaged absolute relative error:
\begin{eqnarray}\label{eqn:optimal_d}
d^*=\arg\min_{d\in\mathcal{D}} \frac{1}{n-(P+1)}\sum_{q'\in \mathcal{I}'\setminus \mathcal{I}'_b}|e_{\textrm{rel}}(\hat{R}_d(\mathcal{S}_{q'}))|.
\end{eqnarray}
\end{algorithm}

{\it Remark:} Algorithm \ref{alg:k_fold_val} applies only if $n>P+1$. Also, the training points $R(\mathcal{S}_{x'_1\cdots x'_P})$'s for all $x'_1\cdots x'_P\in \mathcal{I}'_b$ are never used for validation. Only the $R(\mathcal{S}_{x'_1\cdots x'_P})$ for all $x'_1\cdots x'_P\in \mathcal{I}'\setminus \mathcal{I}'_b$ are used for both training and validation but not in the same round. In contrast, in $K$-fold cross validation, each data point is used as for training during $(K-1)/K$ portion of the rounds and for testing during the other $1/K$ portion of the rounds. This is one difference between the cross validation in this paper and the more widely used $K$-fold cross validation.

After $d^*$ is found from (\ref{eqn:optimal_d}), we plug it into Step 4 of Algorithm \ref{alg:model_fit} and fit all the $n$ training points to the model, and use this trained model to estimate the reach of any other subset. In addition, note that we define the relative error of a validation point in (\ref{eqn:rel_err}) as the absolute estimation error normalized by the gap of the upper and lower bounds of that point. Here, we use the relative error rather than the absolute error, because the reach of each subset may significantly vary but we do not expect any single term to dominate the error sum. 

Finally, let us further interpret the values of $d$ when we compare Algorithm \ref{alg:k_fold_val} against Theorem \ref{theorem:zero_loss}. Using the optimal $d^*$ produced by Algorithm \ref{alg:k_fold_val}, Algorithm \ref{alg:model_fit} cannot achieve zero training error any more. This is not surprising because of the cross validation. Compared to the minimum $d$ in Theorem \ref{theorem:zero_loss} to achieve zero training error, the $d^*$ produced by Algorithm \ref{alg:k_fold_val} is smaller. In other words, after we plug $d^*$ into Algorithm \ref{alg:model_fit} and fit all the $n$ training points to the model, the training error will not be zero. 


\subsection{Error Bar via Cross Validation}\label{sec:error_bar}

Given any unobserved subset $\mathcal{S}_q$ where $q\notin \mathcal{I}'$, after the reach estimation $\hat{R}(\mathcal{S}_q)$ is made by Algorithm \ref{alg:model_fit}, what remains to be shown is how to evaluate the uncertainty of $\hat{R}(\mathcal{S}_q)$. In fact, the $100\%$-confidence interval $[\underline{R}(\mathcal{S}_q), \overline{R}(\mathcal{S}_q)]$ is one conservative answer. Besides it, if we desire a narrower, say, $90\%$-confidence interval, we provide a solution via the following algorithm.
\begin{algorithm}
(Error bar)
\begin{enumerate}
    \item With each $q'\in \mathcal{I}'\setminus \mathcal{I}'_b$ defined in (\ref{eqn:optimal_d}), calculate the relative estimation error $e_{\textrm{rel}}(\hat{R}(\mathcal{S}_{q'}))$ defined in (\ref{eqn:rel_err}). 
    
    \item Calculate the $\alpha^{th}$ quartile, denoted by $q_{\alpha}$, of the absolute relative estimation errors. That is, $q_{\alpha}$ is the $\alpha^{th}$ quartile after sorting the $|e_{\textrm{rel}}(\hat{R}(\mathcal{S}_{q'}))|$'s for all $q'\in \mathcal{I}'\setminus \mathcal{I}'_b$ in an ascending order.
    
    \item For the unobserved subset $\mathcal{S}_q$ where $q\notin \mathcal{I}'$, calculate $\hat{R}(\mathcal{S}_q)$,  $\underline{R}(\mathcal{S}_q)$, $\overline{R}(\mathcal{S}_q)$ by using Algorithm \ref{alg:model_fit} and Theorem \ref{theorem:bound_subset_reach}, respectively. Then for any given the quartile $\alpha$, e.g., $\alpha=90$, the $\alpha\%$-confidence interval is given by $[\underline{R}_{\alpha\%}(\mathcal{S}_q), \overline{R}_{\alpha\%}(\mathcal{S}_q)]$ where
    \begin{eqnarray}
    \underline{R}_{\alpha\%}(\mathcal{S}_q) &\!\!\!\!=\!\!\!\!& \max(\underline{R}(\mathcal{S}_q), ~\hat{R}(\mathcal{S}_q) - q_{\alpha}/2\cdot (\overline{R}(\mathcal{S}_q)-\underline{R}(\mathcal{S}_q)))\\
    \overline{R}_{\alpha\%}(\mathcal{S}_q) &\!\!\!\!=\!\!\!\!& \min(\overline{R}(\mathcal{S}_q),~\hat{R}(\mathcal{S}_q) + q_{\alpha}/2\cdot (\overline{R}(\mathcal{S}_q)-\underline{R}(\mathcal{S}_q))).
    \end{eqnarray}
\end{enumerate}
\end{algorithm}
    
    

{\it Remark:} While the standardized definition of the error bar for our framework is unavailable, we introduce the above algorithm as a start to define and calculate the error bar. Note that the quartile calculation is based on the assumption that the relative estimation error (nearly) follows a normal distribution. If this assumption is avoided, then the definition of the error bar can also be revised to other different forms.

\section{Experiments on Synthetic Data} \label{sec:simulation} 

In this section, we evaluate the performance of the framework proposed in Sec. \ref{sec:framework} through two experiments on synthetic data. In the first experiment, we assume the training points are given and fixed, i.e., without adaptive selection. In the second experiment, we focus on evaluating the performance of adaptively selecting the training points. 

\subsection{Performance of Estimating the Reach of Subsets with Fixed Training Points}

In this experiment, we fix the number of training points as $n=2P+1$. In particular, besides the must-be-included $P$ single-BG reach and the all-BGs union reach, we also include the reach of all-but-one-BGs union subsets as the training points. For $P$ BGs, there are $P$ all-but-one-BGs, i.e., each obtained by excluding one BG from the all the $P$ BGs. Thus, the training points are given by $R(\mathcal{S}_{x'_1\cdots x'_P})$'s for all $x'_1\cdots x'_P\in \mathcal{I}'=\{x'_1\cdots x'_P~|~\sum_{i=1}^P x'_i=k,~k\in\{1,~P-1,~P\}\}$. Furthermore, to mimic the actual reach measurements in presence of the DP noise, we add an independent Gaussian noise to each training point so that the $90^{th}$ quartile of the error on each training point is $10\%$. As a result, the noise variance is $(\frac{0.1}{1.645}R(\mathcal{S}_{x'_1\cdots x'_P}))^2$ for the training point $R(\mathcal{S}_{x'_1\cdots x'_P})$ based on these assumptions. Then we treat the noisy training points as the ground truth in this experiment.

To generate the synthetic data, we employ the following three data generators:
\begin{itemize}
    \item CI(10 groups, random reach): we set 10 groups with the weights drawn from the uniform distribution between 0 and 1. In each group the single-BG reach proportion $r(\mathcal{G}_i)$'s are independently drawn from the Beta(0.4, 2) distribution. 
    
    \item Dirichlet(alpha=2): The reach of the primitive regions is drawn from the Dirichlet distribution with all the entries of the $\alpha$ vector equal to two. 
    
    \item Dirichlet(alpha=0.5): The reach of the primitive regions is drawn from the Dirichlet distribution with all the entries of the $\alpha$ vector equal to $0.5$. In fact, the smaller entry of the alpha vector, the larger variance the reach of the primitive region. Thus, Dirichlet(alpha=0.5) is considered for a stress test.
\end{itemize}
With each data generator, we generate a total of $100$ replicates. Using these 100 replicates in the framework in Figure \ref{fig:framework} allows us to statistically evaluate the reach estimations of the unobserved subsets.

To evaluate the reach estimation of a subset, for each replica of the synthetic data, we first feed the $2P+1$ training points to model fitting in Figure \ref{fig:framework}, including ``Conditional independence model" and ``Tune model parameters via cross validation". Next, using the well-trained model, we estimate the reach of each of the $2^P-1-(2P+1)=2^P-2P-2$ unobserved subsets.
In particular, using the model-free and model-based approaches introduced in Sec. \ref{sec:model_free} and Sec. \ref{sec:DMM}, we calculate the relative error defined in (\ref{eqn:rel_err}) for each estimation. With 100 replicates of the synthetic data, we finally can collect a total of $100(2^P-2P-2)$ error terms. Finally, the $90^{th}$ quartile of these $100(2^P-2P-2)$ error terms, referred to as $q90$(error on reach estimation), can be calculated.

\begin{table}[!ht]
\centering
\caption{Performances of evaluating the framework shown in Figure \ref{fig:framework} on testing }\label{tab:performance_framework}
\vspace{0.5em}
\begin{tabular}{ccc}
    \hline\hline
    $P$ (number of BGs) & Synthetic data & $q90$(error on reach estimations) \\ \hline
    6 & CI(10 groups, random reach) & $10.1\%$ \\ 
    6 & Dirichlet(alpha=2) & $10.9\%$ \\ 
    6 & Dirichlet(alpha=0.5) & $19.8\%$ \\ \hline
    8 & CI(10 groups, random reach) & $10.4\%$ \\ 
    8 & Dirichlet(alpha=2) & $6.4\%$ \\ 
    8 & Dirichlet(alpha=0.5) & $12.7\%$ \\ \hline\hline
\end{tabular}
\end{table}

In Table \ref{tab:performance_framework}, we show $q90$(error on reach estimations) as a performance evaluation metric by assuming $P=6,~8$ BGs for each data generator. It turns out the $90^{th}$ quartiles of the relative error are $10.1\%$ and $10.9\%$ when using CI(10 groups, random reach) and Dirichlet(alpha=2) as the data generator, respectively. That is, $90\%$ of estimations have relative error less than $10.1\%$ and $10.9\%$, respectively. When Dirichlet(alpha=0.5) is used for the purpose of stress test, the performance significantly degrades, but the $90^{th}$ quartiles of the relative error are still below $20\%$. 



\subsection{Performance of Adaptively Selecting the Training Points for Incremental Reach Estimation}

In this section, we revisit the example shown in Figure \ref{fig:incremenral_reach_problem} in Sec. \ref{sec:intro} again. To demonstrate the performance of adaptive training-points selection, we assume there are $P=5$ BGs, each single-BG reach is $100000$, and the universe size $U=500000$. In addition, we assume the probability of a user being reached by one BG is independent with the others. With these assumptions, for any two BGs, any three BGs, any four BGs, and all the five BGs, their union reach can be readily calculated and given by $180000$, $244000$, $295200$, and $336160$, respectively, which align with the numbers used in the example in Figure \ref{fig:incremenral_reach_problem}. In the experiment, we use all these numbers as the ground truth. 

Recall that the training points must include all the single-BG reach and the all-BGs union reach. Thus, we start from using six training points $R(\mathcal{S}_{x'_1\cdots x'_5})=100000$ for any $x'_1\cdots x'_5$ satisfying $\sum_{i=1}^5 x'_i=1$ and $R(\mathcal{S}_{11111})=336160$ to train the model. As our goal is estimate $R(\mathcal{S}_{11000})$, $R(\mathcal{S}_{11100})$, and $R(\mathcal{S}_{11110})$ in the incremental sequence
$G_1\rightarrow G_2\rightarrow G_3\rightarrow G_4\rightarrow G_5$, we call them the testing points. Among the total of $2^P-1=2^5-1=31$ subsets (excluding the subset of users not reached by any BG), other than the 6 training and 3 testing points, the reach of the rest $31-6-3=22$ subsets comprise of the candidates to be measured and used as the additional training points.

\begin{figure}[!t]
\centering
\includegraphics[scale=.50]{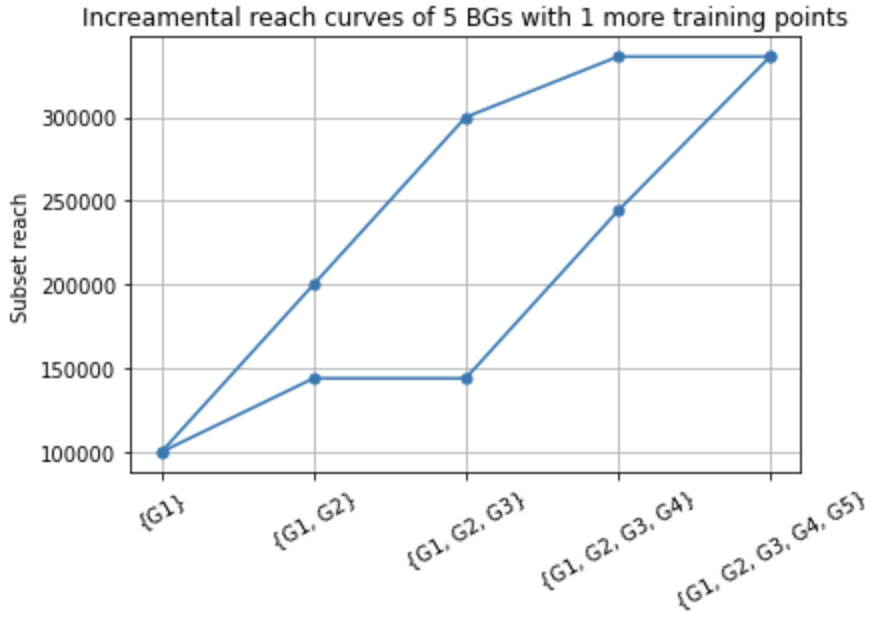}
\includegraphics[scale=.50]{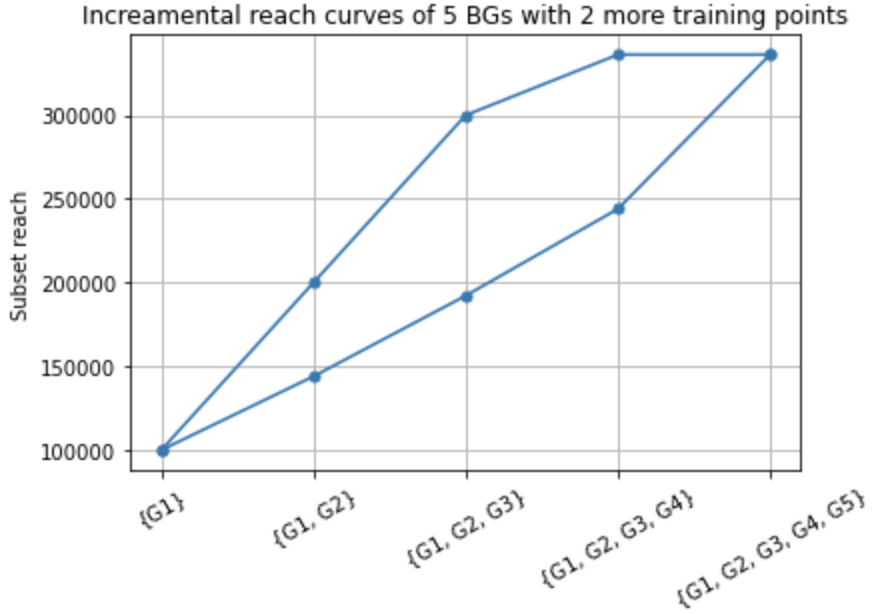}
\\\vspace{-0.1in}
{\small (a) Select 1 more training point. \quad\quad\quad\quad\quad\quad\quad\quad    
(b) Select 2 more training points.}
\\\vspace{0.2in}
\includegraphics[scale=.50]{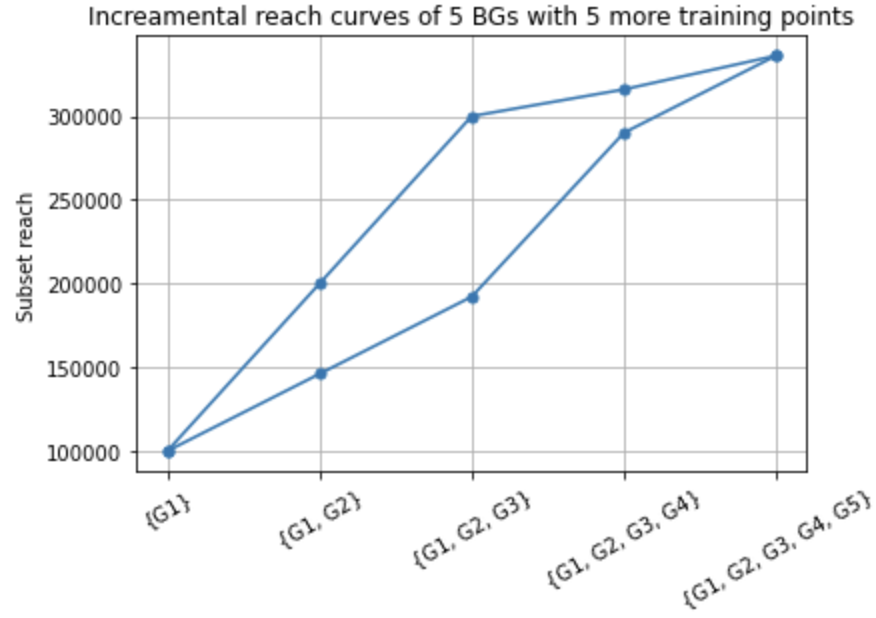}
\includegraphics[scale=.50]{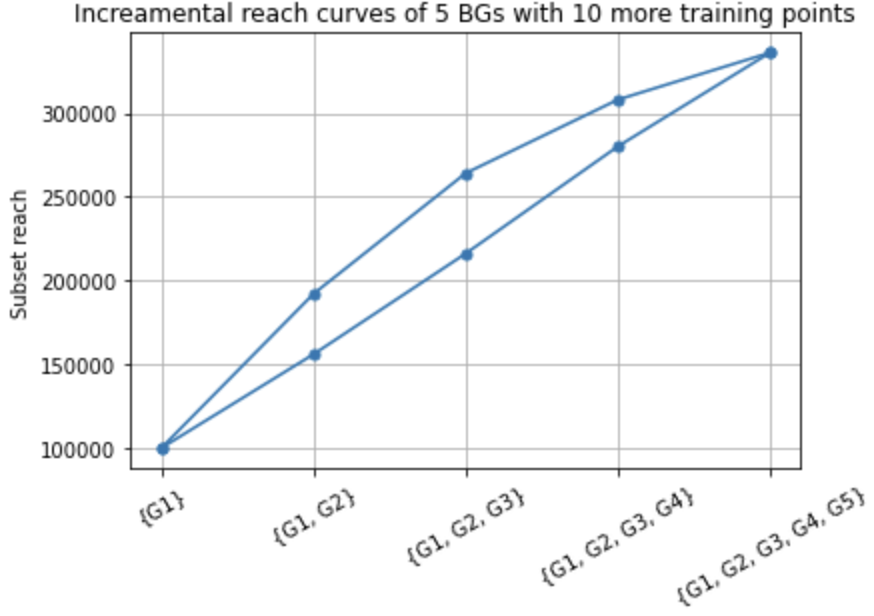}
\\\vspace{-0.1in}
{\small (c) Select 5 more training points.  \quad\quad\quad\quad\quad\quad\quad\quad   
(d) Select 10 more training points.}
\\
\caption{Performance of adaptively selecting up to 10 training points}
\label{fig:perf_adaptive_select}
\end{figure}

In our experiment, we use Algorithm \ref{alg:adaptive_select_train} to select up to 10 additional training points. After each additional training is selected, we use \emph{all} the training points to examine the possible values of the 3 testing points by using the model-free approach introduced in Theorem \ref{theorem:bound_subset_reach}. As a result, the following ten additional training points are selected sequentially in the experiment:
$R(\mathcal{S}_{01011})$, $R(\mathcal{S}_{11010})$, $R(\mathcal{S}_{01110})$, $R(\mathcal{S}_{11100})$, $R(\mathcal{S}_{01101})$, $R(\mathcal{S}_{10110})$, $R(\mathcal{S}_{10011})$, $R(\mathcal{S}_{10101})$, $R(\mathcal{S}_{11001})$, and $R(\mathcal{S}_{00110})$. In Figure \ref{fig:perf_adaptive_select}, we show the range for the reach of the three testing-subsets after we select the first one, first two, first five, and all the ten additional training points shown above. 

Several observations can be made from Figure \ref{fig:perf_adaptive_select}. First, as expected, the more training points, the less uncertainty of the estimations. Second, although selecting more training points is generally useful, we may not always reduce the $100\%$ confidence interval width. For example, the estimation range of $R(\{G_1, G_2\})$ (or $R(\mathcal{G}_1\cup \mathcal{G}_2)$) in sub-figure (b) appears identical to that in sub-figure (a), even though one more training point is used. Finally, let us compare the sub-figure (c) of to Figure \ref{fig:incremenral_reach_linear_programming}. Although five additional training points are used for both, the area enclosed by the blue-colored bounds in sub-figure (c) is larger than that in Figure \ref{fig:incremenral_reach_linear_programming}. This is because we select the subset with the largest reach estimation uncertainty as the next training point, but in principle this may not lead to the best estimations of the incremental reach curves. If we focus on estimating the incremental reach curve only, then we can tailor the algorithm design to directly minimize the area enclosed by the upper and lower bounds of incremental reach curves. Since estimating the incremental reach is only one application of estimating the reach of a subset, we do not investigate the more advanced algorithm in this paper.

\section{Conclusion}\label{sec:conclusion}

In this paper, we investigated the problem of estimating the reach Venn diagram given partial observations of the diagram. A critical use case of this problem is to estimate the incremental reach on any permutations of multiple BGs. To solve the problem, we propose two new approaches.
The first is a model-free approach, meaning that it does not rely on any model assumption. In this approach, we translate the problem of detecting the reach consistency among the reach of the observed subsets into solving a linear-programming problem on the reach Venn diagram. Through imposing constraints of the reach of the observed subsets, we also provide a solution to bound the reach of an unobserved subset. 
The second approach uses the conditional independence model. In particular, we developed a new fitting algorithm for the model that is interpretable. Through solving a non-negative least squares (NNLS) problem, we can provide a point estimate for the reach of any unobserved subset. 
We also provide a framework by integrating these two approaches, combined with training points selection and parameter fine-tuning through cross-validation, to give both confidence interval and point estimates. 

We like to point out that this paper primarily focuses on introducing novel methodologies for estimating the reach of any subset, and the experiments are performed on synthetic data only. 
More comprehensive evaluations will be expected when the real-world data arrives.

\appendix
\section*{Appendix}

\section{The Proof of Theorem \ref{theorem:zero_loss}}

Since it is very challenging to directly prove Theorem \ref{theorem:zero_loss}, we first let $d\rightarrow +\infty$ and show the proof as an intermediate result. Then based on this intermediate result, we complete the proof for Theorem \ref{theorem:zero_loss} in a relatively easier manner.

\subsection{The Intermediate Proof for $d\rightarrow +\infty$}

For each subset $\mathcal{S}_{x'_1x'_2\cdots x'_P}$ defined in Definition \ref{def:minterm_maxterm}, it can be represented as a union of all the mutually-exclusive primitive regions belonging to that subset: 
\begin{eqnarray}
\mathcal{S}_{x'_1x'_2\cdots x'_P}=\cup_{\mathcal{R}_{x_1x_2\cdots x_P}\subseteq \mathcal{S}_{x'_1x'_2\cdots x'_P}} \mathcal{R}_{x_1x_2\cdots x_P},~~~~x'_1x'_2\cdots x'_P \in \mathcal{I}',
\end{eqnarray}
where the primitive region $\mathcal{R}_{x_1x_2\cdots x_P}$ is defined in Definition \ref{def:pr}, and we rewrite it here again: 
\begin{eqnarray}
\mathcal{R}_{x_1\cdots x_P}=\cap_{\forall x_i=1}~\mathcal{G}_i-\cup_{\forall x_i=0}~ \mathcal{G}_i.
\end{eqnarray}
Thus, the reach/ cardinality of the subset $\mathcal{S}_{x'_1x'_2\cdots x'_P}$ equals the sum of the reach/ cardinality of each associated primitive region, i.e., 
\begin{eqnarray}
R(\mathcal{S}_{x'_1x'_2\cdots x'_P})=\sum_{\mathcal{R}_{x_1x_2\cdots x_P}\subseteq \mathcal{S}_{x'_1x'_2\cdots x'_P}} R(\mathcal{R}_{x_1x_2\cdots x_P}),~~~~x'_1x'_2\cdots x'_P \in \mathcal{I}'.
\end{eqnarray}
Normalizing both sides of the $n$ equations above by the universe size $U$, we obtain:
\begin{eqnarray}
r(\mathcal{S}_{x'_1x'_2\cdots x'_P})&\!\!\!\!=\!\!\!\!&\sum_{\mathcal{R}_{x_1x_2\cdots x_P}\subseteq \mathcal{S}_{x'_1x'_2\cdots x'_P}} r(\mathcal{R}_{x_1x_2\cdots x_P})\\
&\!\!\!\!=\!\!\!\!& \sum_{x_1x_2\cdots x_P\in \mathcal{I}}  \mathbbm{1}_{\mathcal{S}_{x'_1x'_2\cdots x'_P}}(\mathcal{R}_{x_1x_2\cdots x_P}) \cdot r(\mathcal{R}_{x_1x_2\cdots x_P}), ~~~~x'_1x'_2\cdots x'_P \in \mathcal{I}',
\end{eqnarray}
where $\mathbbm{1}_{\mathcal{S}}(\mathcal{R})$ is an indicator function where $\mathbbm{1}_{\mathcal{S}}(\mathcal{R})=1$ if $\mathcal{R} \subseteq \mathcal{S}$ and $0$ for otherwise. Stacking the $n$ entities above into a column vector ${\bf r}_{\mathcal{S}_{\mathcal{I}'}}$  by arranging the entries in an ascending order w.r.t. $(\textrm{bin2dec}(x'_P\cdots x'_2x'_1))^{th}$, we rewrite all the $n$ equations above into a compact matrix form:
\begin{eqnarray}\label{eqn:stack_pr2subset}
{\bf r}_{\mathcal{S}_{\mathcal{I}'}} = {\bf U}_{n\times 2^P} \cdot {\bf r}_{\mathcal{R}_{\mathcal{I}}}
\end{eqnarray}
where ${\bf r}_{\mathcal{R}_{\mathcal{I}}}$ is a $2^P \times 1$ column vector with $r(\mathcal{R}_{x_1\cdots x_P})$ as the $(\textrm{bin2dec}(x_P\cdots x_1))^{th}$ entry (index starts from zero) for all $x_1\cdots x_P\in \mathcal{I}$. Moreover, the resulting $n\times 2^P$ matrix ${\bf U}$ describes the probability of any user in the primitive region
$\mathcal{R}_{x_1x_2\cdots x_P}$ being reached by the subset $\mathcal{S}_{x'_1x'_2\cdots x'_P}$. The entry of ${\bf U}$ corresponding to row $\mathcal{S}_{x'_1x'_2\cdots x'_P}$ column $\mathcal{R}_{x_1x_2\cdots x_P}$ can be specified as:
\begin{eqnarray}\label{eqn:matU}
u(\mathcal{S}_{x'_1x'_2\cdots x'_P}, \mathcal{R}_{x_1x_2\cdots x_P})= \mathbbm{1}_{\mathcal{S}_{x'_1x'_2\cdots x'_P}}\mathcal{R}_{x_1x_2\cdots x_P} = \oplus_{\forall i,~x_i=x'_i=1} 1,
\end{eqnarray}
where $\oplus$ denotes the ``logical or" operator.
It can be seen that the value of (\ref{eqn:matU}) is one if there exists a BG $G_i$ so that $x_i=1$ and $x'_i=1$; otherwise, the value is zero. 

Based on (\ref{eqn:stack_pr2subset}), it can be seen that for any observed vector ${\bf r}_{\mathcal{S}_{\mathcal{I}'}}$ that satisfies the reach consistency, there must exist a vector  ${\bf r}_{\mathcal{R}_{\mathcal{I}}}$ with non-negative entries to satisfy ${\bf r}_{\mathcal{S}_{\mathcal{I}'}}={\bf U}\cdot {\bf r}_{\mathcal{R}_{\mathcal{I}}}$. Note that the sum of the entries of the vector ${\bf r}_{\mathcal{R}_{\mathcal{I}}}$ is given by 
\begin{eqnarray}\label{eqn:reach_sum_lessthan1}
\|{\bf r}_{\mathcal{R}_{\mathcal{I}}}\|_1=\sum_{x_1\cdots x_P\in \mathcal{I}} r(\mathcal{R}_{x_1\cdots x_P})=\sum_{x_1\cdots x_P\in \mathcal{I}} R(\mathcal{R}_{x_1\cdots x_P})/U= 1
\end{eqnarray}
due to the fact that the sum of the reach/cardinality of each of the $2^P$ primitive regions equals the universe size.

Next, consider the $n$ training points from the $n$ subsets $\mathcal{S}_{x'_1\cdots x'_P}$ for all $x'_1\cdots x'_P\in \mathcal{I}'$. With Algorithm \ref{alg:model_fit} introduced in Sec. \ref{sec:improved_DMM}, the entry of ${\bf Z}$ can be obtained from (\ref{eqn:z_entry}) and re-stated here:
\begin{eqnarray}
z(\mathcal{S}_{x'_1x'_2\cdots x'_P}, \mathcal{U}_{x_1x_2\cdots x_P})= 1-\Pi_{\forall i,~x'_i=1}(1-r_{x_i}(\mathcal{G}_i)).
\end{eqnarray}
By assuming $d\rightarrow +\infty$, BG $G_i$ has $r_0(\mathcal{G}_i)=r(\mathcal{G}_i)/d=0$ (resp. $r_1(\mathcal{G}_i)=1-(1-r(\mathcal{G}_i))/d)=1$) probability to reach a user in any activity segment $\mathcal{U}_{x_1x_2\cdots x_P}$ with $x_i=0$ (resp. $x_i=1$). Substituting the resulting $r_0(\mathcal{G}_i)=0$ and $r_1(\mathcal{G}_i)=1$ into the above equation, we can further simplify it as:
\begin{eqnarray}\label{eqn:activity_matZ}
z(\mathcal{S}_{x'_1x'_2\cdots x'_P}, \mathcal{R}_{x_1x_2\cdots x_P})=1-\Pi_{\forall i,~x_i=x'_i=1}(1-r_1(\mathcal{G}_i))= \oplus_{\forall i,~x'_i=x_i=1} 1.
\end{eqnarray}
The comparison between (\ref{eqn:matU}) and (\ref{eqn:activity_matZ}) reveals $z(\mathcal{S}_{x'_1x'_2\cdots x'_P}, \mathcal{U}_{x_1x_2\cdots x_P})=u(\mathcal{S}_{x'_1x'_2\cdots x'_P}, \mathcal{R}_{x_1x_2\cdots x_P})$ given the same $x'_1x'_2\cdots x'_P$ and $x_1x_2\cdots x_P$. Thus, we have ${\bf Z}={\bf U}$.

Finally, to prove there exists a $2^P \times 1$ column vector ${\bf w}^*$ with non-negative entries and $\|{\bf w}^*\|_1\leq 1$ so that $J({\bf w}^*)=0$, it suffice to show ${\bf r}_{\mathcal{S}_{\mathcal{I}'}}={\bf Z}\cdot {\bf w}^*$. To see it, we rewrite
\begin{eqnarray}
{\bf r}_{\mathcal{S}_{\mathcal{I}'}}={\bf Z}\cdot {\bf w}^*= {\bf U}\cdot {\bf w}^*.
\end{eqnarray}
Then we directly let ${\bf w}^*={\bf r}_{\mathcal{R}_{\mathcal{I}}}$ in (\ref{eqn:stack_pr2subset}) and thus $\|{\bf w}^*\|_1= \|{\bf r}_{\mathcal{R}_{\mathcal{I}}}\|_1$. The equal sign holds here because the segment $\mathcal{U}_{00\cdots 0}$ and the non-active segment (which includes the users not reached by any BG) together comprise of the primitive region $\mathcal{R}_{00\cdots 0}$. On the other hand, as the non-active segment is not of interest and excluded after Step 1 of Algorithm \ref{alg:model_fit}, we have $\|{\bf w}^*\|_1\leq \|{\bf r}_{\mathcal{R}_{\mathcal{I}}}\|_1$ for any value of $d$. Since the entries of ${\bf r}_{\mathcal{R}_{\mathcal{I}}}$ are non-negative and $\|{\bf r}_{\mathcal{R}_{\mathcal{I}}}\|_1= 1$ owing to (\ref{eqn:reach_sum_lessthan1}), we conclude the entries of ${\bf w}^*$ are also non-negative and $\|{\bf w}^*\|_1\leq \|{\bf r}_{\mathcal{R}_{\mathcal{I}}}\|_1=1$, and the equal sign holds when $d\rightarrow +\infty$. 

So far, we complete the proof for $d\rightarrow +\infty$. \qed

\subsection{The Proof for $d<+\infty$}

In last section with $d\rightarrow +\infty$, $J(\bf w^*)=0$ implies that the observed vector ${\bf r}_{\mathcal{R}_{\mathcal{I}}}$ can be represented by a linear combination of the columns of the matrix ${\bf Z}(+\infty)$, the linear weights are non-negative, and their sum is upper bounded by one. From the geometric perspective, it implies that in a $n$-dimensional vector space where each dimension is for one training point, the $n$-dimensional vector ${\bf r}_{\mathcal{S}_{\mathcal{I}'}}$ must stay in the convex hull of the $2^P$ vertices corresponding to all the columns of ${\bf Z}(+\infty)$, i.e., ${\bf r}_{\mathcal{S}_{\mathcal{I}'}}\in \textrm{Conv}(\textrm{cols}({\bf Z}(+\infty)))\triangleq \mathcal{C}_{\infty}$. 

Next, consider $d$ takes a finite value. Denote by $\mathcal{C}_{d}$ the convex hull of the $2^P$ vertices corresponding to all the columns of ${\bf Z}(d)$. 

First of all, if $d=1$, then both $r_1(\mathcal{G}_i)$ and $r_0(\mathcal{G}_i)$ defined in Definition \ref{def:column_z} would be identical to $r(\mathcal{G}_i)$ after simplification. Without high/low reach probability variations, the $2^P$ activity segments without difference among themselves will be merged into one segment, and the proposed model in Algorithm \ref{alg:model_fit} essentially reduces to an independence model. Meanwhile, the resulting convex hull ${\bf Z}(d)$ collapses into a single point in the $2^P$-dimensional vector space, and this single point cannot be guaranteed to be just the point represented by the observed vector ${\bf r}_{\mathcal{S}_{\mathcal{I}'}}$. Hence, generally we have ${\bf r}_{\mathcal{S}_{\mathcal{I}'}}\notin \mathcal{C}_1$.

Second, consider $1<d<+\infty$. Note that the $2^P$ vertices of $\mathcal{C}_{d}$ are defined by the $2^P$ columns of the matrix ${\bf Z}(d)$, and each of its entry is a continuous function of the parameter $d$. When $d$ increases/decreases, the convex hull $\mathcal{C}_{d}$ also continuously inflates/shrinks. According to the intermediate value theorem, there must exist a finite value $d^*$, so that the observed vector ${\bf r}_{\mathcal{S}_{\mathcal{I}'}}$ just stays on the surface of $\mathcal{C}_{d^*}$ or stays in  $\mathcal{C}_{d^*}$. That is, there always exists ${\bf w}^*$ with non-negative entries and its entry sum upper bounded by one, so that $J_d({\bf w}^*)=0$.

Therefore, we complete all the proof. \qed

{\it Remark:} The proof above can be further reinforced by proving there exists a $d^*$ so that ${\bf r}_{\mathcal{S}_{\mathcal{I}'}}$ stays on the surface of $\mathcal{C}_{d^*}$, and for any $d>d^*$, ${\bf r}_{\mathcal{S}_{\mathcal{I}'}}$ stays in $\mathcal{C}_{d^*}$. The intuition is that for any $1<d_2<d_1<+\infty$, it can be shown that $\mathcal{C}_{1} \subset \mathcal{C}_{d_2} \subset \mathcal{C}_{d_1} \subset \mathcal{C}_{\infty}$, meaning $\mathcal{C}_{d}$ monotonically inflates/shrinks when $d$ increases/decreases. Therefore, there exists not many but only one $d^*$ so that ${\bf r}_{\mathcal{S}_{\mathcal{I}'}}$ stays on the surface of $\mathcal{C}_{d^*}$. To complete this reinforced proof, it suffices to show any point in $\mathcal{C}_{d_2}$ must stay in $\mathcal{C}_{d_1}$ but the vertices of $\mathcal{C}_{d_1}$ stay outside of $\mathcal{C}_{d_2}$. Since this is out of the scope of this paper, we omit the reinforced proof.



\end{document}